\documentclass[onecolumn,twoside]{IEEEtran}
\usepackage[left=0.67in,right=0.67in,top=0.66in,bottom=0.67in]{geometry}
\usepackage[T1]{fontenc}
\usepackage[latin9]{inputenc}
\usepackage{amsthm}
\usepackage{amsmath} 
\usepackage{graphicx} 
\usepackage{color} 
\usepackage{cite}
\usepackage{algpseudocode}
\usepackage{algorithm}
\usepackage{amssymb}
\usepackage{subfigure}
\usepackage{esint}
\usepackage{algpseudocode}
\usepackage{epstopdf}
\usepackage{multirow}
\usepackage{makecell}
\usepackage{float}
\usepackage{lipsum}
\usepackage{balance}
\usepackage{tikz}
\usetikzlibrary{shapes ,arrows, positioning} 
\usepackage{adjustbox}
\usepackage{bbm}
\usepackage{hyperref}
\usepackage{bm}

\newtheorem{definition}{Definition}
\newtheorem*{definition*}{Definition}
\newtheorem{remark}{Remark}

\newtheorem{example}{Example}

\theoremstyle{plain}
\theoremstyle{plain}
\newtheorem{theorem}{Theorem}

\newcommand{\comment}[1]{}

\IEEEoverridecommandlockouts

\usepackage{colortbl}
\usepackage{hhline}

\begin{document}
%
\title{Eliminating Media Noise While Preserving Storage Capacity: Reconfigurable Constrained Codes for Two-Dimensional Magnetic Recording}

\author{
\IEEEauthorblockN{Iven Guzel$^*$,~\IEEEmembership{Student Member,~IEEE},
Do\u{g}ukan \"{O}zbayrak$^*$,
Robert Calderbank,~\IEEEmembership{Life Fellow,~IEEE}, \\ and
Ahmed Hareedy,~\IEEEmembership{Member,~IEEE}}

\thanks{$^*$Iven Guzel and Do\u{g}ukan \"{O}zbayrak have equal contribution to this work.

Do\u{g}ukan \"{O}zbayrak and Ahmed Hareedy are with the Department of Electrical and Electronics Engineering, Middle East Technical University, 06800 Ankara, Turkey (e-mail: dogukan.ozbayrak@metu.edu.tr and ahareedy@metu.edu.tr).

Iven Guzel is with the Department of Electrical and Computer Engineering, University of Illinois Urbana Champaign, Champaign, IL 61820 USA (e-mail: iguzel2@illinois.edu).

Robert Calderbank is with the Department of Electrical and Computer Engineering, Duke University, Durham, NC 27708 USA (e-mail: robert.calderbank@duke.edu).

This work was supported in part by the T\"{U}B\.{I}TAK 2232-B International Fellowship for Early Stage Researchers.}
}


\markboth{}%
{}
%




\IEEEtitleabstractindextext{%
\begin{abstract}

Magnetic recording devices are still competitive in the storage density race with solid-state devices thanks to new technologies such as two-dimensional magnetic recording (TDMR). TDMR offers remarkable storage density increase without the need for new magnetic materials; however, advanced data processing schemes are needed to guarantee reliability. Data patterns where a bit is surrounded by complementary bits at the four positions with Manhattan distance $1$ on the TDMR grid are called plus isolation (PIS) patterns, and they are error-prone. Recently, we introduced lexicographically-ordered constrained (LOCO) codes, namely optimal plus LOCO (OP-LOCO) codes, with minimal redundancy that prevent these patterns from being written in a TDMR device. However, in the high-density regime or the low-energy regime (as the device ages), additional error-prone patterns emerge, specifically data patterns where a bit is surrounded by complementary bits at only three positions with Manhattan distance $1$, and we call them incomplete plus isolation (IPIS) patterns. In this paper, we present capacity-achieving codes that forbid both PIS and IPIS patterns in TDMR systems with wide read heads. Because of their shape, we collectively call the PIS and IPIS patterns rotated T isolation (RTIS) patterns, and we call the new codes optimal T LOCO (OT-LOCO) codes. We analyze OT-LOCO codes and derive their simple encoding-decoding rule that allows reconfigurability. We also present a novel bridging idea for these codes to further increase the rate. Our simulation results demonstrate that OT-LOCO codes not only remarkably outperform OP-LOCO codes, but also entirely eliminate media noise effects at practical TD densities in the range $[0.6, 0.8)$ with high rates in the range $[0.81, 0.83]$. To further preserve the storage capacity, we suggest using OP-LOCO codes, which have higher rates than OT-LOCO codes, early in the device lifetime, then employing the reconfiguration property to switch to OT-LOCO codes later in the device lifetime. While the point of reconfiguration on the density/energy axis is decided manually at the moment, the next step is to use machine learning to make that decision based on the TDMR device status. Moreover, we introduce another coding scheme to remove RTIS patterns in TDMR systems which offers lower complexity, lower error propagation, and track separation, at the expense of a limited rate loss.

\end{abstract}


}

\maketitle

\IEEEdisplaynontitleabstractindextext

%
\IEEEpeerreviewmaketitle

\section{Introduction}\label{sec_intro}

The fierce storage density race between magnetic products and solid-state products has motivated creativity in a variety of areas, such as physics, architecture, and data processing, in order to invent new storage technologies. One of the cutting-edge magnetic technologies is two-dimensional magnetic recording (TDMR) \cite{wood_tdmr, chan_tdmr}. Since its introduction, the TDMR technology has promised storage densities of up to 10 terabits per square inch \cite{wood_tdmr, victora_10tb, seagate}. What makes TDMR specifically attractive is that the additional density increase compared with one-dimensional magnetic recording products emerges through architectural ideas, such as track squeezing and shingled writing \cite{chan_tdmr, mohsen_tdmr}, along with advanced data processing schemes \cite{shayan_tdmr, pituso_tdmr}, i.e., without the need for new magnetic materials.

Constrained codes prevent errors from happening in data storage and transmission systems. Since Shannon discussed constrained systems in 1948 \cite{shan_const}, these codes have found a wide range of applications in various technologies. In early one-dimensional magnetic recording (ODMR) devices, constrained codes were used to control transition separation, remarkably contributing to the density increase \cite{tang_bahl, siegel_mr}, and they are still used to improve reliability in modern ODMR devices \cite{siegel_const, ahh_loco}. In Flash memory devices, constrained codes are used to mitigate inter-cell interference due to charge propagation \cite{lee_ici}, which can extend the device lifetime \cite{veeresh_mlc, ahh_qaloco}. These codes also find application in optical recording devices \cite{immink_opt} as well as multiple data-transmission computer standards \cite{sridhara_ctalk}. A recent study of the power spectral density for some of these codes in data storage systems can be found in \cite{jes_psd}.

The scientific debate on whether to use finite-state machines (FSMs) or lexicographic indexing to design constrained codes started more than 50 years ago. While Franaszek among others designed various FSM-based constrained codes in the 1960s and the 1970s \cite{franaszek}, the first design of run-length-limited (RLL) constrained codes, presented by Tang and Bahl in 1970, was based on lexicographic indexing \cite{tang_bahl}. In 1983, Adler, Coppersmith, and Hassner introduced a systematic method to design FSM-based constrained codes \cite{ach_fsm}, and many researchers adopted their method afterwards \cite{siegel_mr, siegel_const}. While a variety of constrained codes based on lexicographic indexing (also called enumerative codes) were introduced \cite{blake_enum, gu_lex, braun_lex}, including our recent work \cite{ahh_loco, ahh_qaloco, ahh_qaloco}, a systematic method for such design was missing. In 2022, we introduced a general method to design lexicographically-ordered constrained (LOCO) codes \cite{ahh_general} based on the 1973 result of Cover \cite{cover_lex}, and we used this method to design different LOCO codes for various applications \cite{ahh_rr}. LOCO codes are capacity-achieving, systematic, reconfigurable, and offer simplicity of encoding-decoding \cite{ahh_loco, ahh_general}.

In TDMR, data patterns involving a bit surrounded by complementary bits, i.e., isolated, horizontally and vertically are error-prone \cite{mohsen_tdmr, sharov_TCon}. Since the discretized TD channel impulse response is typically $3 \times 3$, grids of such size are typically considered. These isolation patterns can take the shape of a square, where the bit at the center is surrounded by $8$ complementary bits on the $3 \times 3$ grid \cite{bd_tdmr}, and these are called square isolation (SIS) patterns, or can take the shape of a plus sign, where the bit at the center is surrounded by $4$ complementary bits at Manhattan distance $1$ \cite{mohsen_tdmr, sharov_TCon, halevy_TD}, and these are called plus isolation (PIS) patterns \cite{ahh_general}. Research works studying the capacity of TD constrained codes and suggesting TD bit-stuffing techniques include \cite{halevy_TD}, \cite{kato_TCon}, and \cite{siegel_TCon}. When the TDMR system adopts a wide read head, which accesses $3$ adjacent down tracks at the same time \cite{chan_tdmr, shayan_tdmr}, the TD binary constraints can be converted into one-dimensional $8$-ary constraints, allowing systematic constrained (LOCO) coding schemes designed via the general method. In particular, we introduced optimal plus LOCO (OP-LOCO) codes to eliminate PIS patterns \cite{ahh_general}.\footnote{Throughout the paper, optimality is rate-wise (minimal redundancy).}

Our contribution in this paper is threefold:
\begin{enumerate}
\item We show that a new set of error-prone patterns emerges at higher-interference and/or lower-energy stages of the TDMR device lifetime. In particular, these are data patterns where a bit is surrounded by only $3$ complementary bits at Manhattan distance $1$ on the $3 \times 3$ grid, and they are called incomplete PIS (IPIS) patterns. We focus on IPIS patterns where the victim bit is on the middle down track. Collectively, PIS and IPIS patterns have a T shape that can be rotated, and that is why we call them rotated T isolation (RTIS) patterns. Consequently, we design optimal T LOCO (OT-LOCO) codes to eliminate RTIS patterns. OT-LOCO codes incur limited capacity loss compared with OP-LOCO codes, despite adding many new patterns to forbid. We enumerate OT-LOCO codewords, then develop their simple encoding-decoding rule step-by-step mathematically via the aforementioned general method. We also suggest a novel bridging scheme that allows encoding bits within bridging intervals.

\item We demonstrate the effectiveness of the proposed codes via simulation results performed on a practical TDMR model \cite{mohsen_tdmr}. We show that OT-LOCO codes notably outperform OP-LOCO codes at various densities, where the energy per input bit is fixed. At TD densities in the range from $0.6$ to just below $0.8$, we show that OT-LOCO codes can entirely eliminate media noise, i.e., interference, remarkably improving the performance while preserving the TDMR storage capacity by requiring a rate only between $0.81$ and $0.83$. Moreover, we exploit the reconfigurability feature offered by LOCO codes to switch between OP-LOCO and OT-LOCO codes depending on the performance/capacity. In particular, when the device is fresh (low interference or high energy), OP-LOCO codes are used to increase storage capacity, and we reconfigure to OT-LOCO codes as the performance worsens. Reconfiguration decisions are made based on our assessment at the moment, but machine learning can be used to make such decisions instead based on online identification of device status and/or offline channel modeling.

\item We devise a new constrained coding scheme to eliminate RTIS patterns in TDMR systems at lower complexity and lower error propagation, at the expense of a limited rate loss, compared with OT-LOCO codes. We call the new coding scheme simple T LOCO (ST-LOCO) coding scheme, and it comprises leaving the lower down track in each group of $3$ down tracks uncoded and designing a LOCO code for the upper and middle tracks only. This coding scheme also allows lower track data to be passed immediately to the next data processing stage without waiting for data on the other two tracks. We provide the mathematical analysis of the new coding scheme.
\end{enumerate}

The rest of the paper is organized as follows. In Section~\ref{sec_motiv}, we present the practical TDMR system model we are using and motivate the need for new constrained codes. In Section~\ref{sec_otloco}, we define the proposed OT-LOCO codes, enumerate the codewords, and derive the encoding-decoding rule. We also show how to bridge and find the code rate. In Section~\ref{sec_sims}, we present our simulation results, demonstrating performance gains, and suggest how to reconfigure the used LOCO code. In Section~\ref{sec_stloco}, we introduce and analyze the ST-LOCO coding scheme and show its advantages. In Section~\ref{sec_conc}, we conclude the paper and state some future work.

%
%
%
%

\section{TDMR System Model and Motivation}\label{sec_motiv}

In this section, we describe the practical TDMR system we are adopting, we provide error profiles for the uncoded TDMR system that demonstrate the need for new constrained codes in the high-density regime or the low-energy regime (older device), and we compute the capacity of such codes.

Now, we describe the practical TDMR model used in the experiments and define channel parameters. Each bit in the TD system is considered to be a rectangular cell or grid entry characterized by the \textit{track width}~$TW$ and the \textit{bit period}~$BP$, which are defined as the bit length in the cross-track direction (or equivalently, the width of the down track) and the bit length in the down-track direction (or equivalently, the width of the cross track), respectively.

In the read procedure, $3$ adjacent down tracks are read at the same time by the wide read head the TDMR model adopts \cite{chan_tdmr, shayan_tdmr, bd_tdmr}. The TD read-head impulse response duration at half the amplitude in the cross-track and the down-track directions are defined as $PW_{50,\textup{CT}}$ and $PW_{50,\textup{DT}}$, respectively.

The \textit{TD channel density} $D_{\textup{TD}}$ is then \cite{shayan_tdmr}:
\begin{equation}\label{eqn_tddensity}
D_{\textup{TD}} = \frac{PW_{50,\textup{CT}} \times PW_{50,\textup{DT}}}{TW \times BP}.
\end{equation}
Moreover, the \textit{TD energy metric} $E_{\textup{TD}}$ we are using is the area of a bit (a cell) on the TD grid, i.e.,
\begin{equation}\label{eq:TD_energy}
E_{\textup{TD}} = TW \times BP
\end{equation}
Increasing the TD channel density $D_{\textup{TD}}$ in~\eqref{eqn_tddensity} results in degraded system performance as it intensifies interference in both down and cross track directions. Decreasing the TD energy metric $E_{\textup{TD}}$ in~\eqref{eq:TD_energy} results in degraded system performance due to higher impact of media noise.

Let the number of down tracks in the TD grid be $D$, and let the indices of down tracks be $0$, $1$, $2$, $3$, \dots, $D-1$, where $3 \mid D$. Then, considering the wide read head adopted by our TDMR model, the down tracks can be partitioned into groups of $3$ adjacent tracks to be read simultaneously as $(0, 1, 2)$, $(3, 4, 5)$, $(6, 7, 8)$, \dots, $(D-3, D-2, D-1)$. Interference in the cross-track direction from a group into another group is negligible \cite{chan_tdmr, bd_tdmr}. Consequently, we have the discretized TD channel (read-head) impulse response as a $3 \times 3$ matrix representing the intersection of $3$ adjacent down tracks in the same group with $3$ consecutive cross tracks.

\begin{figure}[t]
\centering
    \subfigure[The detrimental square isolation or SIS patterns (shaped as a square).]
    {\label{fig:det_pattern_SIS}
        \begin{tikzpicture}
        \draw[step=1cm] (0,0) grid (3,3);
            \filldraw[fill=lightgray, draw=darkgray] (1.5,1.5) circle (0.35cm);
            \node at (0.5,0.5) {0};
            \node at (1.5,0.5) {0};
            \node at (2.5,0.5) {0};
            \node at (0.5,1.5) {0};
            \node at (1.5,1.5) {1};
            \node at (2.5,1.5) {0};
            \node at (0.5,2.5) {0};
            \node at (1.5,2.5) {0};
            \node at (2.5,2.5) {0};
        \draw[step=1cm] (4,0) grid (7,3);
            \filldraw[fill=lightgray, draw=darkgray] (5.5,1.5) circle (0.35cm);
            \node at (4.5,0.5) {1};
            \node at (5.5,0.5) {1};
            \node at (6.5,0.5) {1};
            \node at (4.5,1.5) {1};
            \node at (5.5,1.5) {0};
            \node at (6.5,1.5) {1};
            \node at (4.5,2.5) {1};
            \node at (5.5,2.5) {1};
            \node at (6.5,2.5) {1};
        \end{tikzpicture}
    } \\
    \subfigure[The detrimental plus isolation or PIS patterns, (shaped as a plus sign).]
    {\label{fig:det_pattern_PIS}
        \begin{tikzpicture}
        \draw[step=1cm] (0,0) grid (3,3);
            \filldraw[fill=lightgray, draw=darkgray] (1.5,1.5) circle (0.35cm);
            \node at (0.5,0.5) {$\cdot$};
            \node at (1.5,0.5) {0};
            \node at (2.5,0.5) {$\cdot$};
            \node at (0.5,1.5) {0};
            \node at (1.5,1.5) {1};
            \node at (2.5,1.5) {0};
            \node at (0.5,2.5) {$\cdot$};
            \node at (1.5,2.5) {0};
            \node at (2.5,2.5) {$\cdot$};
        \draw[step=1cm] (4,0) grid (7,3);
            \filldraw[fill=lightgray, draw=darkgray] (5.5,1.5) circle (0.35cm);
            \node at (4.5,0.5) {$\cdot$};
            \node at (5.5,0.5) {1};
            \node at (6.5,0.5) {$\cdot$};
            \node at (4.5,1.5) {1};
            \node at (5.5,1.5) {0};
            \node at (6.5,1.5) {1};
            \node at (4.5,2.5) {$\cdot$};
            \node at (5.5,2.5) {1};
            \node at (6.5,2.5) {$\cdot$};
        \end{tikzpicture}
    } 
    \caption{The detrimental SIS and PIS patterns. An error is highly likely to occur on the circled bit at the center even if the device is relatively fresh.}
    \label{fig:det_pattern_org}
\end{figure}
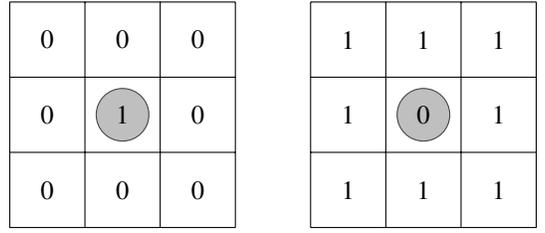
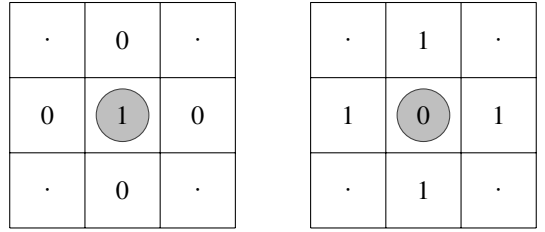

Since level-based signaling is adopted here, a $0$ is converted into level $-A$ indexed by $0$, and a $1$ is converted into level $+A$ indexed by $1$ upon writing. Interference on the same down track (inter-symbol interference or ISI) and on the same cross track (inter-track interference or ITI) can result in the level at the central position of any $3 \times 3$ TD grid changing its sign, which results in an error upon reading if this level is \textit{isolated}. This means the level at the central position is surrounded on the TD grid by $8$ levels, by $4$ levels at Manhattan distance $1$, or by $3$ levels at Manhattan distance $1$ with the complementary sign, and the sets of equivalent $3 \times 3$ binary patterns resulting in such isolation are the set of square isolation (SIS) patterns, the set of plus isolation (PIS) patterns, and the set of incomplete plus isolation (IPIS) patterns, respectively. The set of PIS and IPIS patterns collectively is the set of rotated T isolation (RTIS) patterns. RTIS patterns subsume PIS patterns, and PIS patterns in turn subsume SIS patterns.

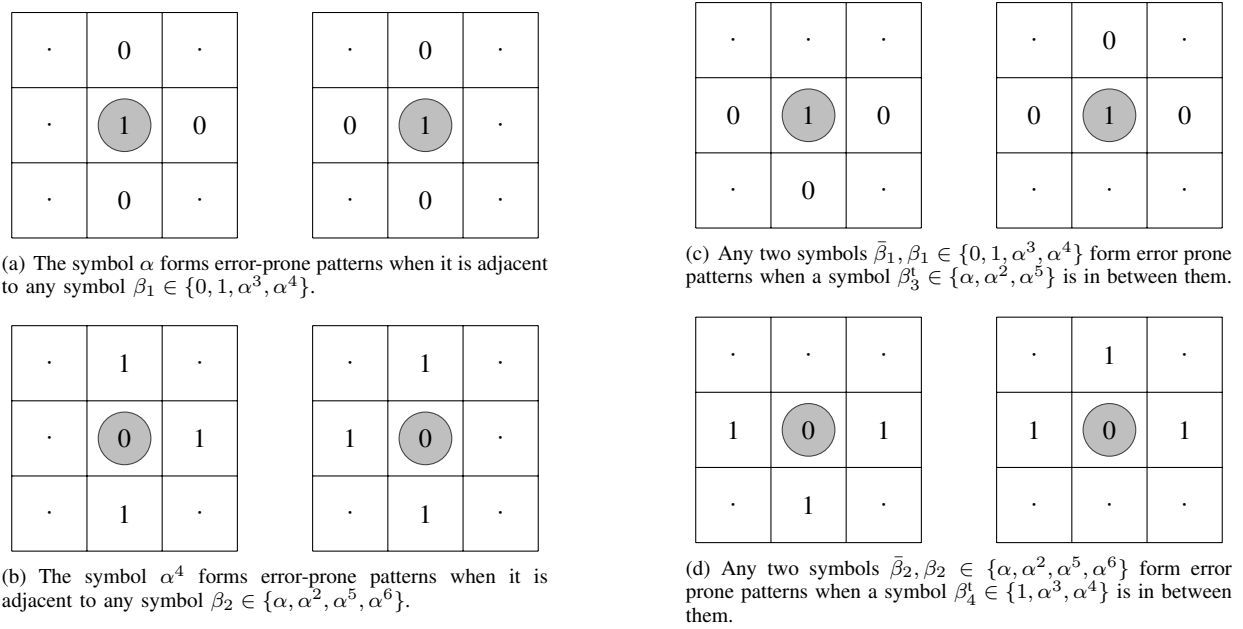
\begin{figure*}[th!]
    \begin{minipage}[t!]{0.5\textwidth}
    \centering
    \subfigure[The symbol $\alpha$ forms error-prone patterns when it is adjacent to any symbol $\beta_1 \in \{0,1,\alpha^3,\alpha^4\}$.]
    {\label{fig:det_pattern_RTIS_a}
        \begin{tikzpicture}
        \draw[step=1cm] (0,0) grid (3,3);
        \filldraw[fill=lightgray, draw=darkgray] (1.5,1.5) circle (0.35cm);
            \node at (0.5,0.5) {$\cdot$};
            \node at (1.5,0.5) {0};
            \node at (2.5,0.5) {$\cdot$};
            \node at (0.5,1.5) {$\cdot$};
            \node at (1.5,1.5) {1};
            \node at (2.5,1.5) {0};
            \node at (0.5,2.5) {$\cdot$};
            \node at (1.5,2.5) {0};
            \node at (2.5,2.5) {$\cdot$};
        \draw[step=1cm] (4,0) grid (7,3);
        \filldraw[fill=lightgray, draw=darkgray] (5.5,1.5) circle (0.35cm);
            \node at (4.5,0.5) {$\cdot$};
            \node at (5.5,0.5) {0};
            \node at (6.5,0.5) {$\cdot$};
            \node at (4.5,1.5) {0};
            \node at (5.5,1.5) {1};
            \node at (6.5,1.5) {$\cdot$};
            \node at (4.5,2.5) {$\cdot$};
            \node at (5.5,2.5) {0};
            \node at (6.5,2.5) {$\cdot$};
        \end{tikzpicture}
    } \\
    \subfigure[The symbol $\alpha^4$ forms error-prone patterns when it is adjacent to any symbol $\beta_2 \in \{\alpha,\alpha^2,\alpha^5,\alpha^6\}$.]
    {\label{fig:det_pattern_RTIS_b}
    \begin{tikzpicture}
        \draw[step=1cm] (0,0) grid (3,3);
        \filldraw[fill=lightgray, draw=darkgray] (1.5,1.5) circle (0.35cm);
            \node at (0.5,0.5) {$\cdot$};
            \node at (1.5,0.5) {1};
            \node at (2.5,0.5) {$\cdot$};
            \node at (0.5,1.5) {$\cdot$};
            \node at (1.5,1.5) {0};
            \node at (2.5,1.5) {1};
            \node at (0.5,2.5) {$\cdot$};
            \node at (1.5,2.5) {1};
            \node at (2.5,2.5) {$\cdot$};
        \draw[step=1cm] (4,0) grid (7,3);
        \filldraw[fill=lightgray, draw=darkgray] (5.5,1.5) circle (0.35cm);
            \node at (4.5,0.5) {$\cdot$};
            \node at (5.5,0.5) {1};
            \node at (6.5,0.5) {$\cdot$};
            \node at (4.5,1.5) {1};
            \node at (5.5,1.5) {0};
            \node at (6.5,1.5) {$\cdot$};
            \node at (4.5,2.5) {$\cdot$};
            \node at (5.5,2.5) {1};
            \node at (6.5,2.5) {$\cdot$};
        \end{tikzpicture}
        
    }
    \end{minipage}%
    \begin{minipage}[t!]{0.5\textwidth}
    \centering
    \subfigure[Any two symbols $\bar \beta_1, \beta_1 \in \{0,1,\alpha^3,\alpha^4\}$ form error prone patterns when a symbol $\beta^\textup{t}_3 \in \{\alpha,\alpha^2,\alpha^5\}$ is in between them.]
    {\label{fig:det_pattern_RTIS_c}
    \begin{tikzpicture}
        \draw[step=1cm] (0,0) grid (3,3);
        \filldraw[fill=lightgray, draw=darkgray] (1.5,1.5) circle (0.35cm);
            \node at (0.5,0.5) {$\cdot$};
            \node at (1.5,0.5) {0};
            \node at (2.5,0.5) {$\cdot$};
            \node at (0.5,1.5) {0};
            \node at (1.5,1.5) {1};
            \node at (2.5,1.5) {0};
            \node at (0.5,2.5) {$\cdot$};
            \node at (1.5,2.5) {$\cdot$};
            \node at (2.5,2.5) {$\cdot$};
        \draw[step=1cm] (4,0) grid (7,3);
        \filldraw[fill=lightgray, draw=darkgray] (5.5,1.5) circle (0.35cm);
            \node at (4.5,0.5) {$\cdot$};
            \node at (5.5,0.5) {$\cdot$};
            \node at (6.5,0.5) {$\cdot$};
            \node at (4.5,1.5) {0};
            \node at (5.5,1.5) {1};
            \node at (6.5,1.5) {0};
            \node at (4.5,2.5) {$\cdot$};
            \node at (5.5,2.5) {0};
            \node at (6.5,2.5) {$\cdot$};
        \end{tikzpicture}
    }
    \subfigure[Any two symbols $\bar \beta_2, \beta_2 \in \{\alpha, \alpha^2,\alpha^5,\alpha^6\}$ form error prone patterns when a symbol $\beta^\textup{t}_4 \in \{1,\alpha^3,\alpha^4\}$ is in between them.]{ \label{fig:det_pattern_RTIS_d}
        \begin{tikzpicture}
        \draw[step=1cm] (0,0) grid (3,3);
        \filldraw[fill=lightgray, draw=darkgray] (1.5,1.5) circle (0.35cm);
            \node at (0.5,0.5) {$\cdot$};
            \node at (1.5,0.5) {1};
            \node at (2.5,0.5) {$\cdot$};
            \node at (0.5,1.5) {1};
            \node at (1.5,1.5) {0};
            \node at (2.5,1.5) {1};
            \node at (0.5,2.5) {$\cdot$};
            \node at (1.5,2.5) {$\cdot$};
            \node at (2.5,2.5) {$\cdot$};
        \draw[step=1cm] (4,0) grid (7,3);
        \filldraw[fill=lightgray, draw=darkgray] (5.5,1.5) circle (0.35cm);
            \node at (4.5,0.5) {$\cdot$};
            \node at (5.5,0.5) {$\cdot$};
            \node at (6.5,0.5) {$\cdot$};
            \node at (4.5,1.5) {1};
            \node at (5.5,1.5) {0};
            \node at (6.5,1.5) {1};
            \node at (4.5,2.5) {$\cdot$};
            \node at (5.5,2.5) {1};
            \node at (6.5,2.5) {$\cdot$};
        \end{tikzpicture}
    }
    \end{minipage}
    \vspace{-0.5em}
    \caption{The detrimental rotated T isolation or RTIS patterns (shaped as the letter~T rotated). The likelihood of an error on the circled bit at the center if an RTIS pattern is an IPIS pattern increases as the device ages.}
    \label{fig:det_pattern_RTIS}
\end{figure*}

SIS, PIS, and RTIS patterns earn their names from their shapes, as shown in Fig.~\ref{fig:det_pattern_SIS}, Fig.~\ref{fig:det_pattern_PIS}, and Fig.~\ref{fig:det_pattern_RTIS}, respectively. Consider the bit at the center of the $3 \times 3$ TD grid. There are only $2$ SIS patterns as shown in Fig.~\ref{fig:det_pattern_SIS}. There are $2 \times 2^4=32$ PIS patterns as shown in Fig.~\ref{fig:det_pattern_PIS} since a ``$\cdot$'' in Fig.~\ref{fig:det_pattern_PIS} or Fig.~\ref{fig:det_pattern_RTIS} means $0$ or $1$. For an IPIS pattern to occur, exactly $3$ bits at Manhattan distance $1$ has to be the complements of the bit at the grid central position as shown in Fig.~\ref{fig:det_pattern_RTIS}. Thus, there are $2 \times \binom{4}{3} \times 2^4=128$ IPIS patterns, which results in a total of $32+128 = 160$ RTIS patterns.

SIS patterns are the most detrimental patterns. However, they are less likely to occur under random writing \cite{bd_tdmr, ahh_general}. PIS patterns were introduced since bits at the corners cause less interference than bits at positions with Manhattan distance $1$ with respect to the center \cite{mohsen_tdmr, sharov_TCon}. PIS patterns are $32/2 = 16$ times more likely to occur compared with SIS patterns under random writing. IPIS patterns are less detrimental than PIS patterns, but they are $128/32 = 4$ times more likely to occur on the middle track in each group of down tracks. We show in this section that IPIS patterns dominate the error profile at high density or low energy, motivating the need for new constrained codes that eliminate RTIS patterns.

Next, we discuss our uncoded TDMR system setup. We have the writing setup, the channel setup, and the reading setup.

\textbf{Writing setup:} We generate random binary data. Before writing to the tracks, level-based signaling is applied, which converts each $0$ into $-1$ and each $1$ into $+1$. Upon writing, these $-1$ and $+1$ values will be updated to values depending on $TW$ and $BP$, i.e., the TD bit energy.

\textbf{Channel setup:} Our baseline channel model is the TDMR model in \cite{mohsen_tdmr}, which is a Voronoi model. Here, we only consider media noise/interference. We modify this model such that it is suitable for a wide read head that reads data from $3$ adjacent down tracks simultaneously. In particular, the upper and lower tracks of each group of $3$ adjacent down tracks have additional protection from interference in the cross track direction \cite{ahh_general}. Thus, the middle down track in each group suffers from the highest level of interference compared with the upper and lower down tracks \cite{chan_tdmr, bd_tdmr}.

In the simulations, we have two sweep setups. First, we sweep the TD channel density $D_{\textup{TD}}$ given in \eqref{eqn_tddensity}. This is performed as follows. The parameters $PW_{50,\textup{CT}}$ and $PW_{50,\textup{DT}}$ are fixed at $20.00$ nm and $14.00$ nm, respectively. The parameter $TW$ is swept between $15.81$ nm and $22.36$ nm, while the parameter $BP$ is swept between $11.07$ nm and $15.65$ nm. We keep the ratio $TW/BP$ the same at all sweep points according to:
\begin{equation}\label{eqn_sameratio}
\frac{TW}{BP} = \frac{PW_{50,\textup{CT}}}{PW_{50,\textup{DT}}} = \frac{10}{7}.
\end{equation}
Thus, and using \eqref{eqn_tddensity}, the TD density $D_{\textup{TD}}$ is swept between $1.60$ and $0.80$. Observe that the range of the TD density simulated could be higher in a TDMR system with equalization, detection, and low-density parity-check (LDPC) coding customized for magnetic recording \cite{shayan_tdmr}.

Second, we sweep the TD bit energy metric $E_{\textup{TD}} = TW \times BP$. This is performed as follows. The parameters $PW_{50,\textup{DT}}$ and $BP$ are both fixed at $7.0$ nm. The parameters $PW_{50,\textup{CT}}$ and $TW$ are both swept between $11.14$ nm and $428.57$ nm. We keep the ratio $PW_{50,\textup{CT}}/TW$ the same at all sweep points according to:
\begin{equation}\label{eqn_sameratio}
\frac{PW_{50,\textup{CT}}}{TW} = \frac{PW_{50,\textup{DT}}}{BP} = 1.00.
\end{equation}
Thus, and using \eqref{eqn_tddensity}, the TD density $D_{\textup{TD}}$ is fixed at $1.00$, while $E_{\textup{TD}}$ is swept between $78.0$ and $3000.0$.

The channel input here is grids of random uncoded data with $3$ rows each after signaling is applied. The channel output is created by applying Voronoi media noise/interference to these input grids. Here, we take the aforementioned protection of the upper and lower tracks in each group of $3$ down tracks into account. The channel effect is equivalent to applying the TD convolution between the input grids and the $3 \times 3$ discretized read-head impulse response with media noise.

\textbf{Reading setup:} For each grid with $3$ rows generated from the channel, hard decision is performed based on the value at each position; if the value is less than or equal to zero, the bit is read as $0$, while if the value is greater than zero, the bit is read as $1$. If the read bit differs from the corresponding one at the relevant input grid, a bit error is counted. An error is characterized according to the input grid corresponding to the $3 \times 3$ grid having the read bit at its center as either PIS error, IPIS error, or random error. Because of adopting a wide read head in the TDMR system, the probability that interference causes an error on the upper or the lower track in each group of $3$ adjacent down tracks is notably lower than that on the middle track in the same group.

\begin{figure}
\vspace{-0.5em}
\center
\includegraphics[trim={0.0in 0.0in 0.0in 0.0in}, width=3.0in]{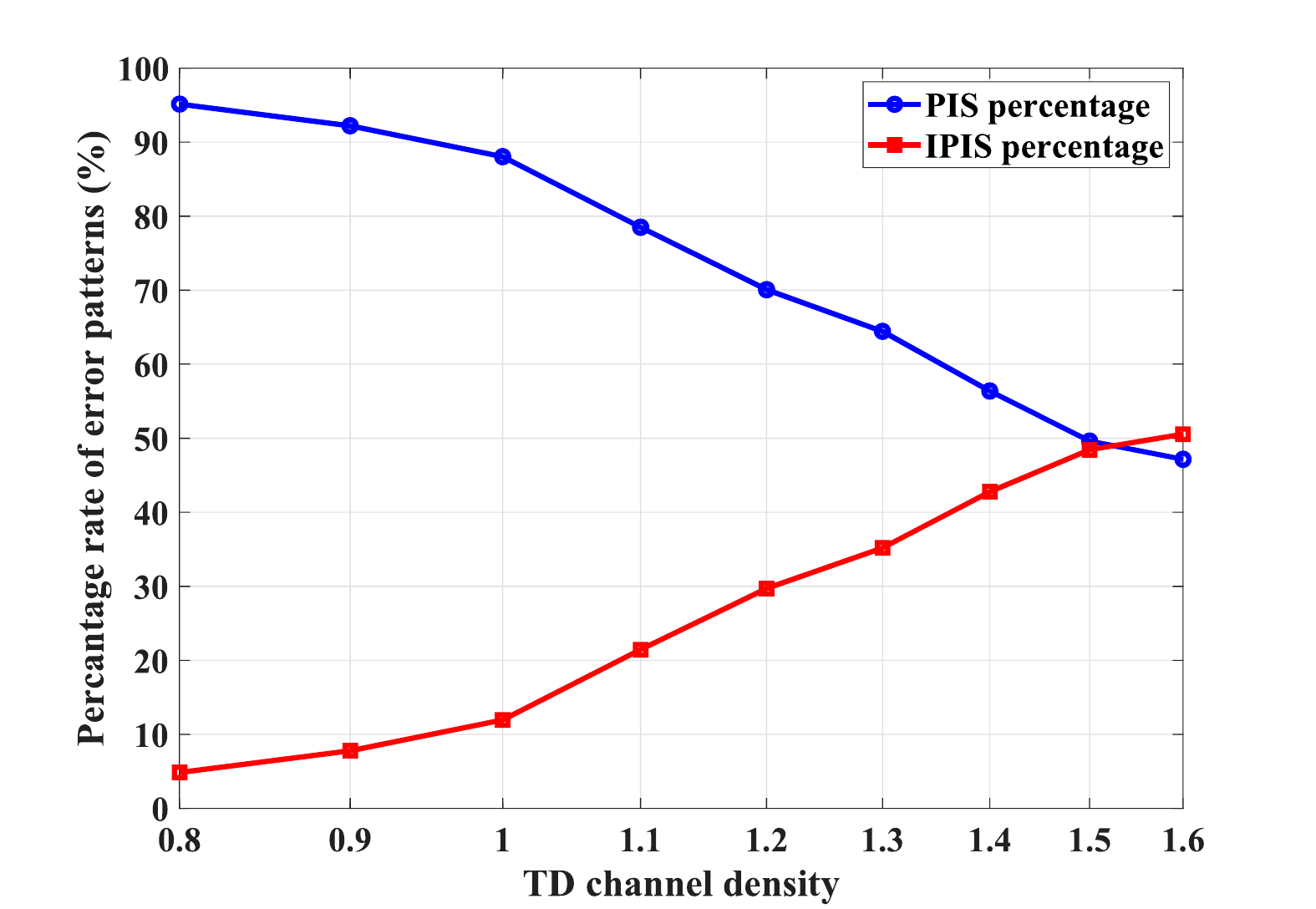}
\vspace{-0.5em}
\caption{The evolution of TDMR error profile with TD density. PIS (IPIS) error percentage decreases (increases) as TD density increases.}
\label{fig_profden}
\vspace{-0.6em}
\end{figure}

\begin{figure}
\vspace{-0.4em}
\center
\includegraphics[trim={0.0in 0.0in 0.0in 0.0in}, width=3.0in]{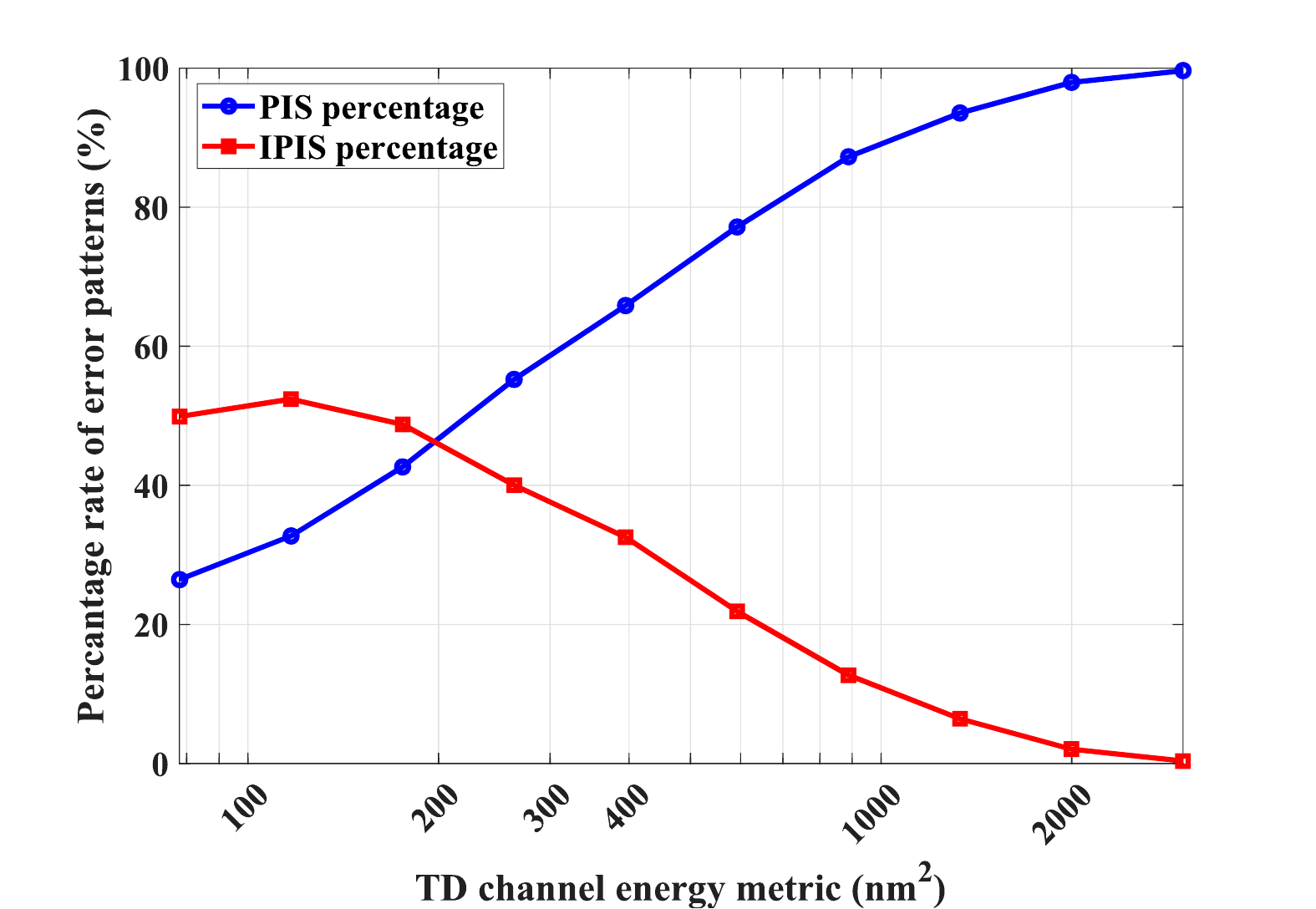}
\vspace{-0.5em}
\caption{The evolution of TDMR error profile with TD energy. PIS (IPIS) error percentage decreases (increases) as TD energy decreases.}
\label{fig_profene}
\vspace{-0.7em}
\end{figure}

We are now ready to discuss the error profiles at the output of the TDMR channel for the uncoded case. Both Fig.~\ref{fig_profden} and Fig.~\ref{fig_profene} confirm the observations that the error profile at low interference levels (limited media noise and fresh device status) is dominated by errors resulting from PIS patterns, or in short PIS errors. In particular, Fig.~\ref{fig_profden} shows that at the low TD density $D_{\textup{TD}}=0.8$, $95.1\%$ of the errors are PIS errors. Moreover, Fig.~\ref{fig_profene} shows that at the high TD energy metric $E_{\textup{TD}}\approx2000$ nm$^2$, $98.0\%$ of the errors are PIS errors. Observe that the lower the TD density or the higher the TD energy, the lower the interference and media noise \cite{shayan_tdmr}. This observation is not surprising, as it is consistent with the findings in, for example, \cite{mohsen_tdmr} and \cite{ahh_general}.

What is unique about these error profile plots we present is they demonstrate that the profile dynamics change as the density increases or as the energy decreases, i.e., as the device gets older. The growth of media noise increases the share of IPIS errors, making it possible for only $3$ complementary bits at Manhattan distance $1$ from the center to cause an error at the center. In particular, Fig.~\ref{fig_profden} shows that at the moderate TD density $D_{\textup{TD}}=1.1$, the share of PIS errors reduces to $78.5\%$, while the share of IPIS errors increases to $21.5\%$. Moreover, Fig.~\ref{fig_profene} shows that at the moderate TD energy metric $E_{\textup{TD}}\approx395$, the share of PIS errors reduces to $65.9\%$, while the share of IPIS errors increases to $32.5\%$. More intriguingly, there is a cross-point on both plots where the share of IPIS errors exceeds that of PIS errors. In particular, Fig.~\ref{fig_profden} shows that at the high TD density $D_{\textup{TD}}=1.6$, the share of PIS errors drops to $47.2\%$, while the share of IPIS errors soars to $50.6\%$. Moreover, Fig.~\ref{fig_profene} shows that at the low TD energy metric $E_{\textup{TD}}=78$, the share of PIS errors drops to $26.4\%$, while the share of IPIS errors soars to $49.9\%$.

Both the density and the energy performance plots we discussed motivate the need for new constrained codes. These new codes should be capable of preventing both PIS and IPIS patterns, and should also be reconfigurable. The reconfigurability property allows us to use constrained codes preventing only PIS patterns, namely OP-LOCO codes, at the early and intermediate stages of the TDMR device lifetime, then switch to constrained codes preventing both PIS and IPIS patterns, namely OT-LOCO codes that prevent collectively RTIS patterns, as the device gets older. The experimental results in this section are the main motivation behind this work.



Systematic efficient TD constrained codes, which are codes that prevent specific data patterns both horizontally and vertically, are notoriously hard to design. They are even quite difficult to study asymptotically \cite{kato_TCon, siegel_TCon}. Here, we make use of a property in the TDMR system, the wide read head as discussed above, to convert the two-dimensional binary constrained coding problem into a one-dimensional non-binary constrained coding problem. This makes the constrained codes we design for TDMR systematic and more efficient than the ones in the literature. Let GF$(2) = \{0,1\}$ and GF$(8) = \{0,1,\alpha,\alpha^2,\dots,\alpha^6\}$, where GF refers to Galois field and $\alpha$ is a primitive element of GF$(8)$. In the new problem, a symbol in GF$(8)$ represents a column with $3$ bits to be written on $3$ adjacent down tracks in the same group. We use the following standard mapping-demapping:
\begin{align}\label{eqn_gf8map}
0 &\longleftrightarrow [ 0 \textup{ } 0 \textup{ } 0 ]^{\textup{T}}, \hspace{+3.6em} 1 \longleftrightarrow [ 0 \textup{ } 0 \textup{ } 1 ]^{\textup{T}}, \nonumber \\
\alpha &\longleftrightarrow [ 0 \textup{ } 1 \textup{ } 0 ]^{\textup{T}}, \hspace{+3.0em} \alpha^2 \longleftrightarrow [ 0\textup{ } 1 \textup{ } 1 ]^{\textup{T}}, \nonumber \\
\alpha^3 &\longleftrightarrow [ 1 \textup{ } 0 \textup{ } 0 ]^{\textup{T}}, \hspace{+3.0em} \alpha^4 \longleftrightarrow [ 1 \textup{ } 0 \textup{ } 1 ]^{\textup{T}}, \nonumber \\
\alpha^5 &\longleftrightarrow [ 1 \textup{ } 1 \textup{ } 0 ]^{\textup{T}}, \hspace{+3.0em} \alpha^6 \longleftrightarrow [ 1 \textup{ } 1 \textup{ } 1 ]^{\textup{T}}.
\end{align}

Next, we mathematically formulate the sets of patterns to forbid in various optimal LOCO codes for TDMR. As discussed above, the reconfigurability feature of LOCO codes can be exploited to prolong the reliability of the TDMR device without trading off storage capacity.

We introduce the sets of $\text{GF}(8)$ patterns equivalent to the SIS, PIS, and RTIS $3 \times 3$ error-prone patterns described above. There are only $2$ SIS patterns, shown in Fig.~\ref{fig:det_pattern_SIS}, which map to the $2$ $\text{GF}(8)$ patterns in the set:
    \begin{equation}
        \mathcal{OS}^8 \triangleq  \{0\alpha0, \ \alpha^6\alpha^4\alpha^6\}. 
        \label{eq:forbid_OS}
    \end{equation}
There are $32$ PIS patterns, shown in Fig.~\ref{fig:det_pattern_PIS}. PIS patterns subsume SIS patterns and map to the $32$ $\text{GF}(8)$ patterns given by the set:
    \begin{align}\label{eq:forbid_OP}
        \mathcal{OP}^8 \triangleq \{\bar\beta_1\alpha\beta_1, \  \bar\beta_2\alpha^4\beta_2, \ \forall \bar \beta_1, \beta_1 \in \{0,1,\alpha^3,\alpha^4\} 
         ,\  \forall \bar\beta_2, \beta_2 \in \{\alpha,\alpha^2,\alpha^5,\alpha^6\}\}.
    \end{align}

In \cite{ahh_general}, optimal square LOCO~(OS-LOCO) and optimal plus LOCO (OP-LOCO) codes, which are optimal with respect to the rate, were proposed to prevent SIS and PIS patterns, respectively. Although SIS patterns are the most detrimental subclass of PIS patterns, it should be noted that the bits at the corners of the $3 \times 3$ grid cause less interference than the bits at positions with Manhattan distance~1 from the center. Also, PIS patterns are $16$ times more likely to occur compared with SIS patterns under unbiased writing. However, preventing PIS patterns for higher reliability incurs some rate loss compared with preventing SIS patterns. Consequently, the normalized capacity $C^n$, for OS-LOCO codes is $C^n =0.9981$~\cite{bd_tdmr}, whereas it is $C^n=0.9710$ for OP-LOCO codes.

In this work and under the reconfigurability setup, we utilize OP-LOCO codes, defined below, to eliminate SIS and PIS patterns at the low-density or the high-energy regime (fresh device).

\begin{definition}[OP-LOCO Code]\label{def:op_loco}
An OP-LOCO code, $\mathcal{OPC}^8_{m}$, is defined by the following properties:
\begin{enumerate}
    \item Codewords in $\mathcal{OPC}^8_{m}$ are defined over $\textup{GF}(8)$, the code alphabet, and are of length $m$ symbols.
    \item Codewords in $\mathcal{OPC}^8_{m}$ are lexicographically ordered.
    \item Codewords in $\mathcal{OPC}^8_{m}$ do not contain any patterns from the set $\mathcal{OP}^8$.
     \item Any codeword satisfying the above properties is included in $\mathcal{OPC}^8_{m}$ .
\end{enumerate}
\end{definition}

A set of sequences defined over GF($q$), where $q$ is the field order/size, is said to be \textit{lexicographically ordered} if its sequences are {sorted in ascending order} following the rule $0 < 1 < \alpha < \alpha^2 < \dots < \alpha^{q-2}$ and the symbol significance gets smaller from left to right.

As per the motivation of the work detailed above, a new set of error-prone patterns emerges as the density increases or the energy decreases (older device), namely the set of IPIS patterns, resulting in a new set of interest, namely the set of RTIS patterns. There are $160$ RTIS patterns ($32$ PIS and $128$ IPIS). While discussing RTIS patterns, we ignore the ``$\cdot$'' bits. The set of RTIS patterns can be viewed as the expansion of $\mathcal{OP}^8$ in \eqref{eq:forbid_OP} into the following four sets of detrimental patterns. $\mathcal{T}_1$ and $\mathcal{T}_2$, illustrated in Fig.~\ref{fig:det_pattern_RTIS_a} and \ref{fig:det_pattern_RTIS_b}, respectively, are the sets of patterns where two of the complementary bits are in the same column as the isolated bit at the center:
    $$\mathcal{T}_1 \triangleq  \{ \alpha\beta_1,\ \beta_1\alpha,\ \forall \beta_1 \in \{0,1,\alpha^3,\alpha^4\}\},$$\vspace{-1.3em}
    $$\mathcal{T}_2 \triangleq  \{ \alpha^4\beta_2,\ \beta_2\alpha^4,\ \forall \beta_2 \in \{\alpha, \alpha^2,\alpha^5,\alpha^6\}\}.$$
$\mathcal{T}_3$ and $\mathcal{T}_4$, illustrated in Fig.~\ref{fig:det_pattern_RTIS_c} and \ref{fig:det_pattern_RTIS_d}, respectively are the sets of patterns where two of the complementary bits are in the same row as the isolated bit at the center (complementary bits surrounding the isolated bit are in $3$ columns):
    $$\mathcal{T}_3 \triangleq  \{ \bar\beta_1'\beta^\textup{t}_3\beta_1', \forall \bar \beta_1, \beta_1 \in \{0,1,\alpha^3,\alpha^4\}, \forall \beta^\textup{t}_3 \in \{\alpha,\alpha^2,\alpha^5\} \},$$\vspace{-1.3em}
    $$\mathcal{T}_4 \triangleq  \{ \bar\beta_2'\beta^\textup{t}_4\beta_2', \forall \bar \beta_2, \beta_2 \in \{\alpha,\alpha^2,\alpha^5,\alpha^6\}, \forall \beta^\textup{t}_4 \in \{1,\alpha,\alpha^3\}\}.$$
We merge these sets of forbidden patterns in a way that the union is \textit{minimal}, i.e., every pattern in the final set of forbidden patterns $\mathcal{OT}^8$ is a \textit{first offender}:
\begin{align}\label{eq:forbid_OT}
    \mathcal{OT}^8 \triangleq \{
    & \alpha\beta_1,\ \beta_1\alpha,\ \alpha^4\beta_2,\ \beta_2\alpha^4,\ \bar\beta_1'\beta_3\beta_1',\   \bar\beta_2'\beta_4\beta_2',
    \ \forall \beta_1 \in \{0,1,\alpha^3,\alpha^4\},\ \forall \bar \beta_1', \beta_1' \in \{0,1,\alpha^3\}, \nonumber \\
    &\ \forall \beta_2 \in \{\alpha,\alpha^2,\alpha^5,\alpha^6\},\ \forall \bar \beta_2', \beta_2' \in \{\alpha^2,\alpha^5,\alpha^6\},
    \ \forall \beta_3 \in \{\alpha^2,\alpha^5\},\ \forall \beta_4 \in \{1,\alpha^3\}
    \}.
\end{align}

Although the set of forbidden patterns $\mathcal{OT}^8$ covers $160$ patterns out of $2^9 = 512$ possible ones for the $3 \times 3$ TDMR grid of bits, as it turns out, the capacity, i.e., the highest achievable rate, of a constrained code preventing these patterns is surprisingly high, which preserves storage capacity. Here, the unique reconfigurability feature of LOCO codes is quite advantageous in prolonging the device lifetime as it enables the same hardware to support multiple LOCO codes. Thus, we introduce new optimal LOCO codes, named optimal T LOCO (OT-LOCO) codes, to improve the reliability of TDMR systems at moderate/long lifetimes by preventing RTIS patterns.

\begin{remark}
     The highest achievable rate of a given $\mathcal{T}$-constrained code $C_m^q$, with alphabet $\textup{GF}(q)$, length $m$, set $\mathcal{T}$ of forbidden patterns, and cardinality $N_q(m)$, is the graph entropy of the finite-state transition diagram (FSTD) representing the constraint. Let the adjacency matrix of the FSTD be $\mathbf{F}$ and the maximum real positive eigenvalue of the characteristic polynomial of $\mathbf{F}$ be $\lambda_{\textup{max}}(\mathbf{F})$. Then, the capacity is \cite{shan_const, siegel_mr}:
     \begin{equation}
         C = \lim_{m \to \infty} \frac{\log_2N_q(m)}{m} = \log_2(\lambda_{\textup{max}}(\mathbf{F})).
         \label{eq:capacity}
     \end{equation}
\end{remark}

\begin{figure}[t]
\centering
\begin{tikzpicture}[->, >= stealth']
    \node [circle, draw] (one) at (-2.5, 0) {$F_1$};
    \node [circle, draw] (two) at (0, 2) {$F_2$};
    \node [circle, draw] (three) at (0, -2) {$F_3$};
    \node [circle, draw] (four) at (2.5, 0) {$F_4$};
    
    \path (one) edge[loop left] node[left,align=center]{$0,1,$\\$\alpha^3$} (one);
    \path (one) edge[bend left] node[left]{$\alpha^2,\alpha^5$} (two);
    \path (one) edge[bend left=13] node[above]{$\alpha^6$} (four);
    \path (one) edge[bend right] node[left]{$\alpha^4$} (three);
    \path (two) edge[loop above] node[above]{$\alpha$} (two);
    \path (two) edge[bend right=15] node[midway,sloped,above]{$\alpha^2,\alpha^5,\alpha^6$} (four);
    \path (four) edge[bend right] node[right]{$\alpha$} (two);
    \path (four) edge[loop right] node[right, align=center]{$\alpha^2,\alpha^5,$\\$\alpha^6$} (four);
    \path (four) edge[bend left] node[right]{$1,\alpha^3$} (three);
    \path (four) edge[bend left=13] node[below]{$0$} (one);
    \path (three) edge[loop below] node[below]{$\alpha^4$} (three);
    \path (three) edge[bend right=12] node[midway,sloped,below]{$0,1,\alpha^3$} (one);
\end{tikzpicture}
\vspace{-0.7em}
\caption{An FSTD representing an infinite $\mathcal{OT}^8$-constrained sequence where patterns in $\mathcal{OT}^8$ are prevented.}\vspace{-0.5em}
\label{fig:FSTD_OT_LOCO}
\end{figure}
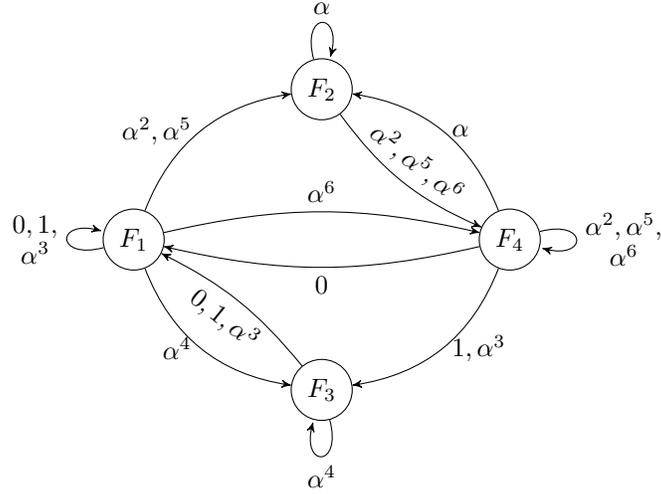

The FSTD of an infinite $\mathcal{OT}^8$-constrained sequence where the patterns in $\mathcal{OT}^8$ are prevented is given in Fig.~\ref{fig:FSTD_OT_LOCO}. 
The corresponding adjacency matrix is:
\begin{equation}
    \mathbf{F}_1 = \begin{bmatrix}
        3 & 2 & 1 & 1 \\ 
        0 & 1 & 0 & 3 \\
        3 & 0 & 1 & 0 \\
        1 & 1 & 2 & 3
        \end{bmatrix}.
    \label{eq:F_OT_LOCO}
\end{equation}
The characteristic polynomial of $\mathbf{F}_1$ is:
\begin{align}
\det(x\mathbf{I}-\mathbf{F}_1) = x^4 - 8x^3 + 15x^2 - 10x - 25 = (x^2 - 3x + 5) (x^2 - 5x - 5).
\end{align}
Using \eqref{eq:capacity}, the capacity $C$, in input bits per coded symbol, and the normalized capacity $C^\textup{n}$ of OT-LOCO codes are:
\begin{align}
    C  &= \log_2(\lambda_{\textup{max}}(\mathbf{F}_1)) = \log_2(5.8541) = 2.5494, \\
    C^\textup{n}  &= \frac{1}{3}C = 0.8498.
\end{align}
Observe that $x=\lambda_{\textup{max}}(\mathbf{F}_1)$ is a root of the irreducible factor $(x^2 - 5x - 5)$ of the characteristic polynomial, which means this factor also verifies the final OT-LOCO cardinality formula \eqref{ot_card}.

\section{Optimal T LOCO (OT-LOCO) Codes}\label{sec_otloco}

In this section, we design the OT-LOCO codes following the general method proposed in~\cite{ahh_general} for designing constrained codes based on lexicographic indexing.
We begin by formally defining OT-LOCO codes:
\begin{definition}[OT-LOCO Code]\label{def:OT_loco}
An OT-LOCO code, $\mathcal{OTC}^8_{m}$, is defined by the following properties:
\begin{enumerate}
    \item Codewords in $\mathcal{OTC}^8_{m}$ are defined over $\textup{GF}(8)$, the code alphabet, and are of length $m$ symbols.
    \item Codewords in $\mathcal{OTC}^8_{m}$ are lexicographically ordered.
    \item Codewords in $\mathcal{OTC}^8_{m}$ do not contain any patterns from the set $\mathcal{OT}^8$ in \eqref{eq:forbid_OT}.
     \item Any codeword satisfying the above properties is included in $\mathcal{OTC}^8_{m}$ .
\end{enumerate}
\end{definition}

Now, we apply the steps of the aforementioned general method to develop the encoding and decoding schemes that are based on the encoding-decoding function $g(\mathbf{c})$, which gives the lexicographic index of any codeword $\mathbf{c}\triangleq c_{m-1}c_{m-2}\dots c_{0}$ in $\mathcal{OTC}^8_{m}$. The main purpose of these steps is to find a formula for the lexicographic index $g(\mathbf{c})$ as a function of codeword symbols and code cardinalities at various lengths.

\textit{First, we specify the group structure.} Using the forbidden patterns in $\mathcal{OT}^8$, we determine the initial groups of $\mathcal{OTC}^8_{m}$ as explained below. Observe that some group merging is applied during the procedure for convenience. Moreover, initial groups of sequences starting with forbidden subsequences according to $\mathcal{OT}^8$ in \eqref{eq:forbid_OT}, e.g., $0\alpha^2\alpha^4$, are directly omitted for brevity.
\begin{itemize}
    \item For the patterns $0\alpha^2\beta'_1$, $\beta'_1 \in \{0,1,\alpha^3\}$, there is an initial group having all the codewords starting with  $0\alpha^2\beta_2$, $\beta_2 \in \{\alpha, \alpha^2,\alpha^5,\alpha^6\}$, from the left. There are six more initial groups having all the codewords starting with $0\beta_5$, a group for each symbol $\beta_5 \in \text{GF}(8) \setminus \{\alpha, \alpha^2\}$, from the left. There are seven more initial groups for codewords starting with each element in $\text{GF}(8) \setminus \{0\}$ from the left. Since all of the initial groups having prefixes of length $1$ are to be eliminated later as there are initial groups with prefixes of length $2$ starting with each element in $\text{GF}(8)$, we skip these initial groups for the rest of the patterns. We do the same for the patterns $1\alpha^2\beta'_1$ and the patterns $\alpha^3\alpha^2\beta'_1$, $\beta'_1 \in \{0,1,\alpha^3\}$.
    
    \item For the patterns $0\alpha^5\beta'_1$, $\beta'_1 \in \{0,1,\alpha^3\}$, there is an initial group having all the codewords starting with  $0\alpha^5\beta_2$, $\beta_2 \in \{\alpha, \alpha^2,\alpha^5,\alpha^6\}$, from the left. There are six more initial groups having all the codewords starting with $0\beta_6$, a group for each symbol $\beta_6 \in \text{GF}(8) \setminus \{\alpha, \alpha^5\}$, from the left. We do the same for the patterns $1\alpha^5\beta'_1$ and the patterns $\alpha^3\alpha^5\beta'_1$, $\beta'_1 \in \{0,1,\alpha^3\}$.
    
    \item Similarly, for the patterns $\alpha^21\beta'_2$ ($\alpha^2\alpha^3\beta'_2$), $\beta'_2 \in \{\alpha^2,\alpha^5,\alpha^6\}$, there is an initial group having all the codewords starting with  $\alpha^21\beta_1$ ($\alpha^2\alpha^3\beta_1$), $\beta_1 \in \{0,1,\alpha^3,\alpha^4\}$, and six more initial groups having all the codewords starting with $\alpha^2\beta_7$ ($\alpha^2\beta_8$), a group for each symbol $\beta_7 \in \text{GF}(8) \setminus \{1,\alpha^4\}$ ($\beta_8 \in \text{GF}(8) \setminus \{\alpha^3,\alpha^4\}$), from the left. We do the same for the patterns $\alpha^51\beta'_2$, ($\alpha^5\alpha^3\beta'_2$) and the patterns $\alpha^61\beta'_2$, ($\alpha^6\alpha^3\beta'_2$), $\beta'_2 \in \{\alpha^2,\alpha^5,\alpha^6\}$.
    
    \item For the patterns $\alpha\beta_1$, there is an initial group having all the codewords starting with $\alpha\beta_2$, $\beta_2 \in \{\alpha, \alpha^2,\alpha^5,\alpha^6\}$, from the left. Similarly, for the patterns, $\alpha^4\beta_2$, there is an initial group having all the codewords starting with $\alpha^4\beta_1$, $\beta_1 \in \{0,1,\alpha^3,\alpha^4\}$, from the left.
    
    \item For the patterns $0\alpha$, $1\alpha$, and $\alpha^3\alpha$, there exist seven initial groups per pattern having all the codewords starting with $0\beta_9$, $1\beta_9$, and $\alpha^3\beta_9$, one for each symbol $\beta_9 \in \text{GF}(8) \setminus \{\alpha\}$, from the left. For the pattern $\alpha^4\alpha$, there is an initial group having all the codewords starting with $\alpha^4\beta_1,\ \beta_1 \in \{0,1,\alpha^3,\alpha^4\}$. Observe that the patterns $\beta_2\alpha$ are complements of the patterns $\beta_1\alpha^4$, and thus, the initial groups for the latter are also complements of the initial groups of the former.
\end{itemize}

Two GF$(q)$ symbols are said to be complements over GF$(q)$ if their integer level-equivalents sum to $q-1$. If $c \in \text{GF}(q)$, its integer level-equivalent is $\mathcal{L}(c) \triangleq \text{gflog}_\alpha(c)+1$ if $c\neq0$, and $\mathcal{L}(c) \triangleq 0$ if $c=0$. Thus, $c_1$ and $c_2$ are complements over GF$(q)$ if $\mathcal{L}(c_1) + \mathcal{L}(c_2) = q-1$. For example, $\alpha$ and $\alpha^4$ are complements over GF$(8)$.

After performing group merging on these initial groups, we end up with $8$ final groups covering all the codewords in $\mathcal{OTC}^8_{m}$:
\begin{itemize}
\item{Group~1} contains all the codewords starting with $0$ from the left.
\item{Group~2} contains all the codewords starting with $1$ from the left.
\item{Group~3} contains all the codewords starting with $\alpha\beta_2$, $\beta_2 \in \{\alpha, \alpha^2,\alpha^5,\alpha^6\}$, from the left.
\item{Group~4} contains all the codewords starting with $\alpha^2$ from the left.
\item{Group~5} contains all the codewords starting with $\alpha^3$ from the left.
\item{Group~6} contains all the codewords starting with $\alpha^4\beta_1$, $\beta_1 \in \{0,1,\alpha^3,\alpha^4\}$, from the left.
\item{Group~7} contains all the codewords starting with $\alpha^5$ from the left. 
\item{Group~8} contains all the codewords starting with $\alpha^6$ from the left.
\end{itemize}

Group~1 is further partitioned into $7$ subgroups:
\begin{itemize}
\item{Subgroup~1.1} contains all the codewords starting with $00$ from the left.
\item{Subgroup~1.2} contains all the codewords starting with $01$ from the left.
\item{Subgroup~1.3} contains all the codewords starting with $0\alpha^2\beta_2$ from the left.
\item{Subgroup~1.4} contains all the codewords starting with $0\alpha^3$ from the left.
\item{Subgroup~1.5} contains all the codewords starting with $0\alpha^4$ from the left.
\item{Subgroup~1.6} contains all the codewords starting with $0\alpha^5\beta_2$ from the left.
\item{Subgroup~1.7} contains all the codewords starting with $0\alpha^6$ from the left.
\end{itemize}
The same partitioning applies to Group~2 and Group~5.\vspace{+0.3em}

Similarly, Group 8 is further partitioned into 7 subgroups:
\begin{itemize}
\item{Subgroup~8.1} contains all the codewords starting with $\alpha^60$ from the left.
\item{Subgroup~8.2} contains all the codewords starting with $\alpha^61\beta_1$ from the left.
\item{Subgroup~8.3} contains all the codewords starting with $\alpha^6\alpha$ from the left.
\item{Subgroup~8.4} contains all the codewords starting with $\alpha^6\alpha^2$ from the left.
\item{Subgroup~8.5} contains all the codewords starting with $\alpha^6\alpha^3\beta_1$ from the left.
\item{Subgroup~8.6} contains all the codewords starting with $\alpha^6\alpha^5$ from the left.
\item{Subgroup~8.7} contains all the codewords starting with $\alpha^6\alpha^6$ from the left.
\end{itemize}
The same partitioning applies to Group~4 and Group~7.

\textit{Second, we enumerate the codewords.} Now, we use the final group structure determined in the first step to derive the cardinality (size) of the code $\mathcal{OTC}^8_{m}$, which is $N_8(m)$. Theorem~\ref{thr:step_2} gives the cardinality of an OT-LOCO code.

\begin{theorem} \label{thr:step_2}
    The cardinality $N_8(m)$ of an OT-LOCO code $\mathcal{OTC}^8_{m}$ is given by:
    \begin{equation}
        N_8(m) = 5N_8(m-1) + 5N_8(m-2), \textup{ } m \ge 2,
        \label{ot_card}
    \end{equation}
    where the defined cardinalities are $N_8(1) \triangleq 8$ and $N_8(0) \triangleq 2$.

\begin{proof}
    Let the cardinality of a codeword group indexed by $i$ in $\mathcal{OTC}^8_{m}$ be $N_{8,i}(m)$. Before we derive the formulae for $N_{8,i}(m),\ i =1,2,\dots,8$, notice the symmetry between certain groups by observing their subgroup structures. Group~3 and Group~6 are symmetric, which means the cardinalities of~these groups are the same. Similarly, the remaining groups are symmetric and they share the same group cardinality. Thus,
        \begin{align}
            N_{8,3}(m) &= N_{8,6}(m), \label{eq:2.1}\\ 
            N_{8,i}(m) &= N_{8,1}(m), \textup{ } i = 2,4,5,7,8. \label{eq:2.2}
        \end{align}
     Consequently, the cardinality of $\mathcal{OTC}^8_{m}$ is:
         \begin{equation}\label{eq:2.3}
             N_8(m) = 6 N_{8,1}(m) + 2N_{8,3}(m).
         \end{equation}
     Moreover, combined cardinalities of the groups starting with $\beta_1 \in \{0,1,\alpha^3,\alpha^4\}$ and the groups starting with $\beta_2 \in \{\alpha, \alpha^2,\alpha^5,\alpha^6\}$ from the left are the same, and each equals half of the code cardinality due to symmetry:
         \begin{equation} \label{eq:2.4}
            \sum_{i \in \{1,2,5,6\}}N_{8,i}(m) = \sum_{i \in \{3,4,7,8\}}N_{8,i}(m) = \frac{1}{2}N_8(m).
         \end{equation}
    
    As for Group~1, notice that concatenating $0$ to a codeword in $\mathcal{OTC}^8_{m-1}$ from the left causes forbidden patterns if the left most symbol (LMS) or symbols of the codeword in $\mathcal{OTC}^8_{m-1}$ are $\alpha$ or $\beta_3\beta_1$, where  $\beta_1 \in \{0,1,\alpha^3,\alpha^4\}, \ \beta_3 \in \{\alpha^2,\alpha^5\}$.
    Thus, each codeword in $\mathcal{OTC}^8_{m}$ starting with $0\gamma$, where $\gamma \in \text{GF}(8) \setminus\{\alpha, \alpha^2, \alpha^5\}$, from the left corresponds to a codeword in $\mathcal{OTC}^8_{m-1}$ starting with the same symbol $\gamma$ and they share the $m-2$ right most symbols (RMSs). This relation is bijective. 
    Moreover, each codeword in $\mathcal{OTC}^8_{m}$ starting with $0\alpha^2\beta_2$ or $0\alpha^5\beta_2$, where $\beta_2 \in \{\alpha, \alpha^2,\alpha^5,\alpha^6\}$, from the left corresponds to a codeword in $\mathcal{OTC}^8_{m-2}$ starting with the same $\beta_2$ and they share the $m-3$ RMSs. 
    This relation is bijective as well.
    Here, we have $\gamma \in \{0,1,\alpha^3,\alpha^4, \alpha^6\}$ and $\beta_2 \in \{\alpha, \alpha^2,\alpha^5,\alpha^6\}$. Thus, the cardinality of Group~1 is:
        \begin{equation}\label{eq:2.5}
            N_{8,1}(m) = \sum_{i \in \{1,2,5,6,8\}}\hspace{-0.3em}N_{8,i}(m-1) \  + \sum_{i \in \{3,4,7,8\}}\hspace{-0.3em}2N_{8,i}(m-2).  
        \end{equation}
    
    As for Group~6, each codeword in $\mathcal{OTC}^8_{m}$ starting with $\alpha^4\beta_1$ is related to a codeword in $\mathcal{OTC}^8_{m-1}$ starting with the same $\beta_1$ from the left and sharing the $m-2$ RMSs.
    Due to the symmetry and using \eqref{eq:2.4}, the cardinality of Group~6 is
        \begin{equation}\label{eq:2.7}
            N_{8,6}(m)  = N_{8,3}(m) = \dfrac{1}{2}N_{8}(m-1),
        \end{equation}
    Utilizing the group symmetries via \eqref{eq:2.1}, \eqref{eq:2.2}, and \eqref{eq:2.4}, Group~1 cardinality in \eqref{eq:2.5} becomes:
        \begin{equation}\label{eq:2.6}
            N_{8,1}(m) = 4N_{8,1}(m-1) + N_{8,3}(m-1) + N_{8}(m-2).
        \end{equation}
    Plugging  \eqref{eq:2.7} into \eqref{eq:2.6} yields:
        \begin{equation}\label{eq:2.8}
            N_{8,1}(m) = 4N_{8,1}(m-1) + \frac{3}{2}N_{8}(m-2).
        \end{equation}
    
    The only remaining step before finding the code cardinality using \eqref{eq:2.3} is to find $N_{8,1}(m-1)$. Plugging \eqref{eq:2.7} into \eqref{eq:2.3} gives the recursive formula for the Group~1 cardinality in terms of code cardinalities:
        \begin{equation}\label{eq:2.9}
            N_{8,1}(m)  = \dfrac{1}{6 }\left(N_{8}(m) - N_{8}(m-1)\right). 
        \end{equation}
    Thus, substituting \eqref{eq:2.7}, \eqref{eq:2.8}, and \eqref{eq:2.9} in \eqref{eq:2.3} gives:
        \begin{align}
            N_8(m) &= 6\left(4N_{8,1}(m-1) + \frac{3}{2}N_{8}(m-2)\right) + N_8(m-1) \nonumber \\
            &= 5N_8(m-1) + 5N_8(m-2). \label{eq:2.10}
        \end{align}
        
    As for the defined cardinalities, it is clear that $N_8(1) \triangleq 8$. Furthermore, once the $14$ forbidden sequences of length $2$ in $\mathcal{OT}^8$ are removed from the $8^2$ possible $2$-tuple $\text{GF}(8)$ sequences, we get $N_8(2) = 50$. Next, $N_8(0)$ is derived from the recursive formula in \eqref{ot_card} as follows:
    \begin{equation}
        N_8(0) = \frac{N_8(2)}{5} - N_8(1) = 2.
    \end{equation}
\end{proof}
\end{theorem}

\vspace{-0.3em}
\textit{Third, we determine the special cases.} The contribution of a symbol $c_i$ in codeword $\mathbf{c} \triangleq c_{m-1}c_{m-2}\dots c_{0}$ in $\mathcal{OTC}^8_{m}$ to the overall codeword index $g(\mathbf{c})$ depends on $c_i$ and its preceding symbols.  
For given a codeword $\mathbf{c} = c_{m-1}c_{m-2}\dots c_{i+1}c_{i}c_{i-1} \dots c_{0}$, the symbol $c_i$ undertakes a \textit{special case} if there exists a codeword $c_{i}'c_{i-1} \dots c_{0}$ in $\mathcal{OTC}^8_{i+1}$ with $c_i' < c_i$ that forms a forbidden pattern when concatenated from the right to $c_{m-1}c_{m-2}\dots c_{i+1}$.
Thus, the contribution of $c_i$ to the codeword index $g(\mathbf{c})$ should be separately calculated. 
On the other hand, a \textit{typical case} of existence for $c_i$ means all codewords starting with any $c_i' < c_i$ from the left in $\mathcal{OTC}^8_{i+1}$ are allowed to be concatenated from the right to the symbols preceding $c_i$.

We now specify the initial special cases for the OT-LOCO code using the patterns in $\mathcal{OT}^8$. Note that some special case merging is also performed here as it is more convenient.
\begin{itemize}
    \item For the patterns $\bar\beta_1'\alpha^2\beta_1',\ \bar \beta_1', \beta_1' \in \{0,1,\alpha^3\}$, two initial special cases of length $3$ are $c_{i+2}c_{i+1}c_{i}=\bar\beta_1'\alpha^2\alpha$ and $c_{i+2}c_{i+1}c_{i}=\bar\beta_1'\alpha^2\alpha^2$. Note that
    $c_{i+2}c_{i+1}c_{i}=\bar\beta_1'\alpha^2\alpha^3$ and $c_{i+2}c_{i+1}c_{i}=\bar\beta_1'\alpha^2\alpha^4$ are eliminated since they are forbidden patterns.
    Finally, $c_{i+2}c_{i+1}c_{i}=\bar\beta_1'\alpha^2\gamma_1,\ \gamma_1 \in \{\alpha^5,\alpha^6\}$ form another initial special case.
    Although $\alpha$ and $\alpha^2$ are consecutive symbols in $\text{GF}(8)$, the first two of these initial special cases are not processed together. The reason is that while calculating the contribution of $c_i=\alpha$ and $c_i=\alpha^2$, symbols $c_i' \in \{0,1\}$ and $c_i' \in \{0,1,\alpha\}$ are considered, respectively.
    Here, we refer the reader to final group descriptions in the first step, where 
    cardinalities of groups consisting of codewords starting with $\alpha$~(Group~3) and $\alpha^4$~(Group~6) are different from the rest of the group cardinalities.
    Other initial special cases are $c_{i+1}c_{i}=\bar\beta_1'\gamma_2,\  \gamma_2 \in \{\alpha^3,\alpha^4\}$, $c_{i+1}c_{i}=\bar\beta_1'\alpha^5$, and $c_{i+1}c_{i}=\bar\beta_1'\alpha^6$. Similarly, these initial special cases are treated separately considering which subgroups starting with $c_{i+1}c'_{i}$, $c'_i <c_i$, are relevant. This processing logic is going to be further illustrated in the fourth step.
    
    \item For the patterns $\bar\beta_1'\alpha^5\beta_1'$, initial special cases of length $3$ are $\bar\beta_1'\alpha^5\alpha$, $\bar\beta_1'\alpha^5\alpha^2$, and $\bar\beta_1'\alpha^5\gamma_1,\ \gamma_1 \in \{\alpha^5,\alpha^6\}$. Moreover, an initial special case of length $2$ is $c_{i+1}c_{i}=\bar\beta_1'\alpha^6$.
    
    \item For the patterns $\bar\beta_2'1\beta_2',\ \bar \beta_2', \beta_2' \in \{\alpha^2, \alpha^5, \alpha^6\}$, the first special case is $c_{i+2}c_{i+1}c_{i}=\bar\beta_2'1\gamma_2,\  \gamma_2 \in \{\alpha^3,\alpha^4\}$.
    Initial special cases of length $2$ are $c_{i+1}c_{i}=\bar\beta_2'\alpha$, $c_{i+1}c_{i}=\bar\beta_2'\gamma_3,\  \gamma_3 \in \{\alpha^2,\alpha^3\}$, and $c_{i+1}c_{i}=\bar\beta_2'\gamma_1,\ \gamma_1 \in \{\alpha^5,\alpha^6\}$.
    
    \item For the patterns $\bar\beta_2'\alpha^3\beta_2'$, there are only two initial special cases where $c_{i+2}c_{i+1}c_{i}=\bar\beta_2'\alpha^3\gamma_2,\  \gamma_2 \in \{\alpha^3,\alpha^4\}$ and $c_{i+1}c_{i}=\bar\beta_2'\gamma_1,\ \gamma_1 \in \{\alpha^5,\alpha^6\}$.
    
    \item For the patterns $\alpha\beta_1,\ \beta_1 \in \{0,1,\alpha^3,\alpha^4\}$, there are three initial special cases: $c_{i+1}c_{i}= \alpha\alpha$, $c_{i+1}c_{i}= \alpha\alpha^2$, and $c_{i+1}c_{i}= \alpha\gamma_1,\ \gamma_1 \in \{\alpha^5,\alpha^6\}$.
    
    \item For the patterns $\beta_1\alpha,\ \beta_1 \in \{0,1,\alpha^3,\alpha^4\}$, we first investigate the patterns $\beta_1'\alpha$, $\beta_1' \in \{0,1,\alpha^3\}$.
    There is an initial special case that is  $c_{i+1}c_{i}= \beta_1'\alpha^2$. Other initial special cases starting with $\beta_1'$ are $c_{i+1}c_{i}= \beta_1'\gamma_2,\  \gamma_2 \in \{\alpha^3,\alpha^4\}$, $c_{i+1}c_{i}= \beta_1'\alpha^5$, and $c_{i+1}c_{i}= \beta_1'\alpha^6$. There is another initial special case for the pattern $\alpha^4\alpha$, which is $c_{i+1}c_{i}= \alpha^4\gamma_2,\  \gamma_2 \in \{\alpha^3,\alpha^4\}$.
    
    \item  For the patterns $\alpha^4\beta_2,\ \beta_2 \in \{\alpha, \alpha^2, \alpha^5, \alpha^6\}$, the only initial special case is  $c_{i+1}c_{i}= \alpha^4\gamma_2,\  \gamma_2 \in \{\alpha^3,\alpha^4\}$.
    
    \item For the patterns $\beta_2\alpha^4$, we investigate the pattern $\alpha\alpha^4$ first.
There is an initial special case that is $c_{i+1}c_{i}= \alpha\gamma_1,\ \gamma_1 \in \{\alpha^5,\alpha^6\}$.
    There is another initial special case for patterns $\beta'_2\alpha^4, \beta_2' \in \{\alpha^2, \alpha^5, \alpha^6\}$, which is $c_{i+1}c_{i}= \beta'_2\gamma_1,\ \gamma_1 \in \{\alpha^5,\alpha^6\}$.
\end{itemize}

\begin{table}[t]
\caption{Final Special Cases for OT-LOCO Codes Grouped According to the Left-Most Symbol(s).}
\centering
\begin{tabular}{|l|l|l|l|} 
\hline
{\cellcolor[rgb]{0.753,0.753,0.753}} \textbf{1}&  \makecell[l]{ $\beta_1'\beta_3\alpha$\\ $\beta_1'\beta_3\alpha^2$\\ $\beta_1'\beta_3\gamma_1,\ \gamma_1 \in \{\alpha^5,\alpha^6\}$} & {\cellcolor[rgb]{0.753,0.753,0.753}}\textbf{4} & \makecell[l]{$\beta_2'\alpha$ \\$\beta_2'\gamma_3,\ \gamma_3 \in \{\alpha^2,\alpha^3\}$\\ $\beta_2'\gamma_1,\ \gamma_1 \in \{\alpha^5,\alpha^6\}$ }  \\ 
\hline
{\cellcolor[rgb]{0.753,0.753,0.753}} \textbf{2}& $\beta_2'\beta_4\gamma_2,\ \gamma_2 \in \{\alpha^3,\alpha^4\}$ & {\cellcolor[rgb]{0.753,0.753,0.753}}\textbf{5} & \makecell[l]{$\alpha\alpha$ \\ $\alpha\alpha^2$\\ $\alpha\gamma_1,\ \gamma_1 \in \{\alpha^5,\alpha^6\}$}  \\ 
\hline
{\cellcolor[rgb]{0.753,0.753,0.753}} \textbf{3}& \makecell[l]{$\beta_1'\alpha^2$\\ $\beta_1'\gamma_2,\ \gamma_2 \in \{\alpha^3,\alpha^4\}$\\ $\beta_1'\alpha^5$ \\$\beta_1'\alpha^6$} & {\cellcolor[rgb]{0.753,0.753,0.753}}\textbf{6}& \makecell[l]{$\alpha^4\gamma_2,\ \gamma_2 \in \{\alpha^3,\alpha^4\}$}  \\ 
\hline
\end{tabular}
\label{tab:final_special_cases}
\end{table}

Denote the total number of final cases by $n_\textup{c}$ and the index of the $i^{th}$ final case by $i_\textup{c}$, $1 \leq i_\textup{c} \leq n_\textup{c}$. After further processing, we end up with $n_\textup{c} =16$ final cases, including the typical case, for $c_i$ based on $c_i$ and its preceding symbols. The $15$ final special cases are further lumped together in $6$ groups based on the similarity between their left most symbols in Table~\ref{tab:final_special_cases}. To simplify the notation, we drop the overline ($\bar\beta'$) in the final cases, as shown in Table~\ref{tab:final_special_cases}, and in the following analysis.

\textit{Fourth, we derive the symbol contribution.}
Denote the contribution of each symbol $c_i$ of $\mathbf{c}\triangleq c_{m-1}c_{m-2}\dots c_{i+1}c_{i} \dots c_{0}$ to the LOCO codeword index $g(\mathbf{c})$ by $g_i(c_i)$. 
According to Cover in~\cite{cover_lex}, $g_i(c_i)$ is the number of  codewords in $\mathcal{OTC}^8_{m}$ starting with the same $m-(i+1)$ symbols of $\mathbf{c}$ from the left and having $c'_i <c_i$ at the $i^{th}$ index, i.e., the number of codewords starting with
$c_{m-1}c_{m-2}\dots c_{i+1}c_{i}'$, for all $c'_i <c_i$, according to the lexicographic ordering definition. 
Equivalently, according to the general method of constructing LOCO codes, $g_i(c_i)$ is the number of codewords in $\mathcal{OTC}^8_{i+1}$ that start with $c_i' < c_i$ and can be concatenated from the right to $c_{m-1}c_{m-2}\dots c_{i+1}$ without forming any of the forbidden patterns in $\mathcal{OT}^8$. 

Let us denote the contribution of $c_i$ in the case indexed by $i_\textup{c}$ by $g_{i,i_\textup{c}}(c_i)$. Then, $g_i(c_i) = g_{i,i_\textup{c}}(c_i)$ for one of the final cases indexed by $1 \leq i_\textup{c} \leq n_\textup{c}$, which is satisfied by $c_i$ and its preceding symbols. Moreover, $g_{i,i_\textup{c}}(c_i)$ can be expressed as a linear combination of cardinalities of OT-LOCO codes having lengths at most $i+1$.

Let $a_i \triangleq \mathcal{L}(c_i)$,  where $\mathcal{L}(c)$ is the integer level-equivalent of a symbol in $\text{GF}(8)$. Recall that if $c \in \text{GF}(q)$, $\mathcal{L}(c) \triangleq \text{gflog}_\alpha(c)+1$ if $c\neq0$, and $\mathcal{L}(c) \triangleq 0$ if $c=0$. Note that the contribution of $c_i = 0$ is $g_{i}(0) = 0$ in all cases since $0$ is the first symbol in the lexicographic order. We start off with the typical case, indexed by $i_\textup{c}=1$. The symbol contribution $g_i(c_i)$ in this case is the number of codewords in $\mathcal{OTC}^8_{m}$ starting with $c_{m-1}c_{m-2}\dots c_{i+1}c_{i}'$ such that $c'_i <c_i$. Equivalently, $g_i(c_i)$ in this case is the number of all codewords in $\mathcal{OTC}^8_{i+1}$ starting with $c'_i$, $c'_i <c_i$, since the typical case is the unrestricted case. Thus, $g_{i,1}(c_i)$ can be calculated as follows:
\begin{itemize}
    \item If $c_i \in \{1,\alpha\}$,
        \begin{align}
            g_{i,1}(c_i) & = \sum_{j=1}^{a_i} N_{8,1}(i+1) \nonumber \\
            & = \frac{a_i}{6}\left( N_8(i+1) - N_8(i)\right), \label{eq:sc1_2}
        \end{align}
    where the second equality in \eqref{eq:sc1_2} follows from \eqref{eq:2.9}.
    \item If $c_i \in \{\alpha^2,\alpha^3,\alpha^4\}$,
        \begin{align}
            g_{i,1}(c_i) & = \sum_{j=1}^{a_i-1} N_{8,1}(i+1) \  + N_{8,3}(i+1) \nonumber \\
            & = \frac{a_i-1}{6}\left( N_8(i+1) - N_8(i)\right) + \frac{1}{2}N_8(i), \label{eq:sc1_4}
        \end{align}
    where the second term in \eqref{eq:sc1_4} follows from \eqref{eq:2.7}.
    \item If $c_i \in \{\alpha^5,\alpha^6\}$,
        \begin{align}
            g_{i,1}(c_i) & = \sum_{j=1}^{a_i-2} N_{8,1}(i+1) \  + 2N_{8,3}(i+1) \nonumber \\
            & = \frac{a_i-2}{6}\left( N_8(i+1) - N_8(i)\right) + N_8(i). \label{eq:sc1_6}
        \end{align}
\end{itemize}

Now, we analyze the first group of the special cases in Table~\ref{tab:final_special_cases}, characterized by $c_{i+2}c_{i+1}=\beta_1'\beta_3, \beta_1' \in \{0,1,\alpha^3\}, \beta_3 \in \{\alpha^2,\alpha^5\}$. The special cases in this group are indexed by $i_\textup{c} \in \{2,3,4\}$ (ordering is always as shown in Table~\ref{tab:final_special_cases}):

\begin{itemize}
    \item If $c_{i+2}c_{i+1}c_i=\beta_1'\beta_3\alpha$, the contribution of $c_i$ to $g(\mathbf{c})$ in this case is the number of codewords in $\mathcal{OTC}^8_{m}$ starting with $c_{m-1}c_{m-2}\dots c_{i+3}\beta_1'\beta_3c_{i}'$ from the left such that $c'_i <c_i=\alpha$. This number equals the number of codewords in $\mathcal{OTC}^8_{i+1}$ starting with $c'_i <c_i=\alpha$ that are allowed to be concatenated to  $c_{m-1}c_{m-2}\dots c_{i+3}\beta_1'\beta_3$ from the right. 
    Thus, the symbol contribution of $c_i$ in this case is 0 since $\beta_1'\beta_3c'_i,\ c'_i  \in \{0,1\}$ is a forbidden pattern:
        \begin{equation}
            g_{i,2}(c_i) = 0. \label{eq:sc2}
        \end{equation}
    \item If $c_{i+2}c_{i+1}c_i=\beta_1'\beta_3\alpha^2$, the contribution of $c_i$ to $g(\mathbf{c})$ in this case is the number of codewords in $\mathcal{OTC}^8_{i+1}$ starting with $c'_i <c_i=\alpha^2$ such that $c'_i  \notin \{0,1\}$ because only $c'_i =\alpha$ that can be concatenated to  $c_{m-1}c_{m-2}\dots c_{i+3}\beta_1'\beta_3$ from the right. Consequently, and using \eqref{eq:2.7}, we derive $g_{i,3}(c_i)$ as follows:
        \begin{equation}
            g_{i,3}(c_i) = N_{8,3}(i+1) = \frac{1}{2}N_8(i). \label{eq:sc3}
        \end{equation}
     \item For the case characterized by $c_{i+2}c_{i+1}c_i=\beta_1'\beta_3\gamma_1, \ \gamma_1 \in \{\alpha^5, \alpha^6\}$, when $c_i = \alpha^5$, only $c'_i  \in \{\alpha,\alpha^2\}$, $c'_i < c_i=\alpha^5$, are possible starting symbols for the codewords in $\mathcal{OTC}^8_{i+1}$ that are allowed to be concatenated to $c_{m-1}c_{m-2}\dots c_{i+3}\beta_1'\beta_3$ from the right. Similarly, when  $c_i = \alpha^6$, possible symbols for  $c'_i <c_i=\alpha^6$ are $c'_i  \in \{\alpha,\alpha^2, \alpha^5\}$. Therefore, $g_{i,4}(c_i)$ is:
        \begin{equation}\label{eq:sc4_0}
            g_{i,4}(c_i)=\begin{cases}
                    N_{8,1}(i+1) + N_{8,3}(i+1), &\text{if} \ c_i = \alpha^5,\\
                    2N_{8,1}(i+1) + N_{8,3}(i+1), &\text{if} \ c_i = \alpha^6.
            \end{cases}
        \end{equation}
    Equivalently, and through using \eqref{eq:2.7} and \eqref{eq:2.9}, we can derive $g_{i,4}(c_i)$ as follows:
        \begin{equation}\label{eq:sc4}
            g_{i,4}(c_i) = \frac{a_i-5}{6}\left( N_8(i+1) - N_8(i)\right) + \frac{1}{2}N_8(i).
        \end{equation}
\end{itemize}

Next, we proceed with the special case indexed by $i_\textup{c} = 5$ and characterized by $c_{i+2}c_{i+1}c_i=\beta_2'\beta_4\gamma_2, \beta_2' \in \{\alpha^2,\alpha^5,\alpha^6\},\  \beta_4 \in \{1,\alpha^3\},\  \gamma_2 \in \{\alpha^3,\alpha^4\}$. The contribution of $c_i$ to $g(\mathbf{c})$ in this case is the number of codewords in $\mathcal{OTC}^8_{m}$ starting with $c_{m-1}c_{m-2}\dots c_{i+3}\beta_2'\beta_4c_{i}'$ from the left such that $c'_i <c_i=\gamma_2$. This number equals the number of codewords in $\mathcal{OTC}^8_{i+1}$ starting with $c'_i <c_i$ and $c'_i \in \{0,1, \alpha^3\}$ because $\beta_2'\beta_4c'_i$ for $c'_i \in \{\alpha,\alpha^2\}$ are forbidden.
Following the same steps leading to \eqref{eq:sc4_0} and \eqref{eq:sc4} for $c_i = \gamma_2 \in \{\alpha^3,\alpha^4\}$ gives:
    \begin{equation}\label{eq:sc5_0}
        g_{i,5}(c_i)=\begin{cases}
            2N_{8,1}(i+1), &\text{if} \ c_i = \alpha^3,\\
            3N_{8,1}(i+1), &\text{if} \ c_i = \alpha^4.
        \end{cases}
    \end{equation}
Thus, $g_{i,5}(c_i)$ is:
    \begin{equation}\label{eq:sc5}
        g_{i,5}(c_i) = \frac{a_i-2}{6}\left( N_8(i+1) - N_8(i)\right).
    \end{equation}
    
Now, we analyze the third group of the special cases in Table~\ref{tab:final_special_cases}, characterized by $c_{i+1}=\beta_1'$, $\beta_1' \in \{0,1,\alpha^3\}$. The special cases in this group are indexed by $i_\textup{c} \in \{6,7,8,9\}$:
\begin{itemize}
    \item If $c_{i+1}c_i=\beta_1'\alpha^2$, the contribution of $c_i$ to $g(\mathbf{c})$ in this case is the number of codewords in $\mathcal{OTC}^8_{i+1}$ starting with $c'_i <c_i$ except for those starting with $\alpha$ from the left. Therefore, $g_{i,6}(c_i)$ can be derived as follows ($a_i=3$):
         \begin{equation}\label{eq:sc6}
             g_{i,6}(c_i) = \hspace{-0.4em} \sum_{j=1}^{a_i-1} N_{8,1}(i+1)= \frac{a_i-1}{6}\left( N_8(i+1) - N_8(i)\right).
         \end{equation}
    \item If $c_{i+1}c_i=\beta_1'\gamma_2, \ \gamma_2 \in \{\alpha^3, \alpha^4\}$, the contribution of $c_i$ to $g(\mathbf{c})$ in this case is the number of codewords in $\mathcal{OTC}^8_{m}$ starting with $c_{m-1}c_{m-2}\dots c_{i+2}\beta_1'c_{i}'$ from the left such that $c'_i <c_i =\gamma_2$. This number is also the number of codewords in $\mathcal{OTC}^8_{i+1}$ starting with $c'_i <c_i$ except for those starting with $\alpha$ or $\alpha^2\beta_1$ from the left. Consequently, and via the symmetry in \eqref{eq:2.4}, $g_{i,7}(c_i)$ is:
            \begin{align}\label{eq:sc7}
                g_{i,7}(c_i) &= \sum_{j=1}^{a_i-2} N_{8,1}(i+1)\  + \frac{1}{2}N_8(i) \nonumber \\
                            &=  \frac{a_i-2}{6}\left( N_8(i+1) - N_8(i)\right) + \frac{1}{2}N_8(i).
            \end{align}
    \item If $c_{i+1}c_i=\beta_1'\alpha^5$, the contribution of $c_i$ to $g(\mathbf{c})$ in this case is the number of codewords in $\mathcal{OTC}^8_{i+1}$ starting with $c'_i <c_i$ except for those starting with $\alpha$ or $\alpha^2\beta_1$ from the left. Thus, we compute $g_{i,8}(c_i)$ in \eqref{eq:sc8_2} by utilizing \eqref{eq:2.4} and \eqref{eq:2.7}:
            \begin{align}
                g_{i,8}(c_i) &= \sum_{j=1}^{a_i-3} N_{8,1}(i+1)\  + \frac{1}{2}N_8(i) + N_{8,3}(i+1) \nonumber \\
                            &=  \frac{a_i-3}{6}\left( N_8(i+1) - N_8(i)\right) + N_8(i). \label{eq:sc8_2}
            \end{align}
    \item Similarly, if $c_{i+1}c_i=\beta_1'\alpha^6$, the contribution of $c_i$ to $g(\mathbf{c})$ in this case is the number of codewords in $\mathcal{OTC}^8_{i+1}$ starting with $c'_i <c_i$ except for those starting with $\alpha$ and $\beta_3\beta_1$, $\beta_3 \in \{\alpha^2,\alpha^5\}$, from the left:
            \begin{align}\label{eq:sc9}
                g_{i,9}(c_i) &= \sum_{j=1}^{a_i-4} N_{8,1}(i+1)\  + N_8(i) + N_{8,3}(i+1) \nonumber \\
                            &=  \frac{a_i-4}{6}\left( N_8(i+1) - N_8(i)\right) + \frac{3}{2}N_8(i).
            \end{align}
\end{itemize}

Observe that in cases where $c_i$ is a specific GF($8$) symbol, $a_i$ is known. However, we still write some equations in terms of $a_i$ for the ease of merging the contributions in the fifth step.

Next, we study the fourth group of special cases in Table~\ref{tab:final_special_cases}, characterized by $c_{i+1}=\beta_2'$, $\beta_2' \in \{\alpha^2,\alpha^5,\alpha^6\}$. Special cases in this group are indexed by $i_\textup{c} \in \{10,11,12\}$ (ordered as in Table~\ref{tab:final_special_cases}). Symbol contributions $g_{i,10}(c_i),\  g_{i,11}(c_i),\  g_{i,12}(c_i)$ are derived following an approach similar to the one we applied for the third group of special cases:
\begin{itemize}
    \item If $c_{i+1}c_i=\beta_2'\alpha$, the symbol contribution in this case $g_{i,10}(c_i)$ equals the number of codewords in $\mathcal{OTC}^8_{i+1}$ starting with $c'_i <c_i$ except for those starting with $1\beta_2$ from the left:
            \begin{align}\label{eq:sc10}
                \hspace{-0.5em}g_{i,10}(c_i) &= \sum_{j=1}^{a_i-1} N_{8,1}(i+1)\  + \frac{1}{2}N_8(i) \nonumber \\
                            &=  \frac{a_i-1}{6}\left( N_8(i+1) - N_8(i)\right) + \frac{1}{2}N_8(i).
            \end{align}
    \item If $c_{i+1}c_i=\beta_2'\gamma_3, \gamma_3 \in \{\alpha^2,\alpha^3\}$, the symbol contribution in this case $g_{i,11}(c_i)$ equals the number of codewords in $\mathcal{OTC}^8_{i+1}$ starting with $c'_i <c_i$ except for those starting with $1\beta_2$ from the left:
            \begin{align}\label{eq:sc11}
                g_{i,11}(c_i) &= \sum_{j=1}^{a_i-2} N_{8,1}(i+1)\  + \frac{1}{2}N_8(i) + N_{8,3}(i+1) \nonumber \\
                            &=  \frac{a_i-2}{6}\left(N_8(i+1) - N_8(i)\right) + N_8(i).
            \end{align}
    \item If $c_{i+1}c_i=\beta_2'\gamma_1, \gamma_1 \in \{\alpha^5, \alpha^6\}$, the symbol contribution in this case $g_{i,12}(c_i)$ equals the number of codewords in $\mathcal{OTC}^8_{i+1}$ starting with $c'_i <c_i$ except for those starting with $\beta_4\beta_2$, $\beta_4 \in \{1,\alpha^3\}$, or $\alpha^4$ from the left:
            \begin{align}\label{eq:sc12}
                g_{i,12}(c_i) &= \sum_{j=1}^{a_i-4} N_{8,1}(i+1)\  + N_8(i) + N_{8,3}(i+1) \nonumber \\
                            &=  \frac{a_i-4}{6}\left(N_8(i+1) - N_8(i)\right) + \frac{3}{2}N_8(i).
            \end{align}
\end{itemize}

As for the fifth group of special cases in Table~\ref{tab:final_special_cases}, which is characterized by $c_{i+1}=\alpha$ and indexed by $i_\textup{c} \in \{13,14,15\}$:
\begin{itemize}
    \item If $c_{i+1}c_i=\alpha\alpha$, none of the codewords in $\mathcal{OTC}^8_{i+1}$ starting with $c'_i <c_i$ can be concatenated to $c_{i+1}=\alpha$ from the right. Hence, $g_{i,13}(c_i)$ is:
        \begin{equation}
            g_{i,13}(c_i) =0. \label{eq:sp13}
        \end{equation}
    \item If $c_{i+1}c_i=\alpha\alpha^2$, $g_{i,14}(c_i)$ is the number of codewords in $\mathcal{OTC}^8_{i+1}$ starting with $\alpha$. Thus,
        \begin{equation}\label{eq:sp14}
            g_{i,14}(c_i) = N_{8,3}(i+1)= \frac{1}{2}N_8(i).
        \end{equation}
    \item If $c_{i+1}c_i=\alpha\gamma_1,\ \gamma_1 \in \{\alpha^5,\alpha^6\}$, $g_{i,15}(c_i)$ is the number of codewords in $\mathcal{OTC}^8_{i+1}$ starting with $c'_i <c_i$ such that $c'_i\in \{\alpha,\alpha^2,\alpha^5\}$ from the left:
            \begin{align}\label{eq:sc15}
                g_{i,15}(c_i) &= \sum_{j=1}^{a_i-5} N_{8,1}(i+1)\ + N_{8,3}(i+1) \nonumber \\
                            &=  \frac{a_i-5}{6}\left(N_8(i+1) - N_8(i)\right) + \frac{1}{2}N_8(i).
            \end{align}
\end{itemize}

Finally, for the special case characterized by $c_{i+1}c_i=\alpha^4\gamma_2, \gamma_2 \in \{\alpha^3,\alpha^4\}$, $g_{i,16}(c_i)$ is the number of codewords in $\mathcal{OTC}^8_{i+1}$ starting with $c'_i <c_i$ except for those starting with $\alpha$ or $\alpha^2$ from the left:
        \begin{align}\label{eq:sc16}
            g_{i,16}(c_i) &= \sum_{j=1}^{a_i-2} N_{8,1}(i+1) \nonumber \\
                        &=  \frac{a_i-2}{6}\left(N_8(i+1) - N_8(i)\right).
        \end{align}

\textit{Fifth, we formulate the encoding-decoding rule.} The lexicographic index $g(\mathbf{c})$ of the codeword  $\mathbf{c} \in \mathcal{OTC}^8_{m}$ can be obtained by the sum of symbol contributions of all symbols constituting $\mathbf{c}$:
    \begin{equation}\label{eq:g_of_c}
        g(\mathbf{c}) = \sum_{i=0}^{m-1} g_i(c_i).
    \end{equation}
Equation \eqref{eq:g_of_c} requires the different expressions of $g_{i,i_\textup{c}}(c_i)$, $1 \leq i_\textup{c} \leq 16$, to be \textit{merged} into a single expression of $g_i(c_i)$. To this end, we define \textit{merging variables} $y_{i,1},\ y_{i,2}, \dots, y_{i,n_\textup{y}}$ such that $n_\textup{y}\leq n_\textup{c}$ for the symbol $c_i$ to switch on the contribution of a specific case out of $n_\textup{c}$ cases from Steps 3 and 4. Moreover, we define \textit{merging functions}, $f_\ell^{\textup{mer}}(\cdot)$, to relate the cardinality $N_8((i+1)-\ell)$ to $g_i(c_i)$ in the unified expression. Finally, we note that $f_\ell^{\textup{mer}}(\cdot)$ has to be function of the merging variables for $c_i$ and $a_i = \mathcal{L}(c_i)$ for the bijection between the codeword and its index  $g(\mathbf{c})$ to be preserved. 
Theorem \ref{thr:step_5} gives the encoding-decoding rule of an OT-LOCO code $\mathcal{OTC}^8_{m}$. The proof of this theorem is also the fifth step of the general method. 

\begin{theorem}\label{thr:step_5}
    Let $\mathbf{c}$ be an OT-LOCO codeword in $\mathcal{OTC}^8_{m}$. The relation between the lexicographic index $g(\mathbf{c})$ of this codeword and the codeword itself is given by:      
         \begin{align}\label{eqn_ruleot}
            g(\mathbf{c}) = \sum_{i=0}^{m-1} \left[ \left( \frac{a_i-\theta_{i,1}}{6}\right) N_8(i+1) + \left( \frac{\theta_{i,1} +3\theta_{i,2} -a_i}{6}\right) N_8(i) \right],
        \end{align}
    where $\theta_{i,1} = 4y_{i,1} +2y_{i,2}+ y_{i,3}$, $\theta_{i,2} = 2y_{i,4}+ y_{i,5}$, and the vector $\mathbf{y}_i$ of merging variables $y_{i,1},\ y_{i,2},\ y_{i,3},\ y_{i,4},\ y_{i,5}$ such that $\mathbf{y}_i=[y_{i,1}\ y_{i,2}\ y_{i,3}\ y_{i,4}\ y_{i,5}]$ is specified as follows:\\
        \indent $\mathbf{y}_i=[01000]$ if $c_{i+2}c_{i+1}c_i=\beta_1'\beta_3\alpha$  $(i_\textup{c}=2)$, else,\\
        \indent $\mathbf{y}_i=[01101]$ if $c_{i+2}c_{i+1}c_i=\beta_1'\beta_3\alpha^2$  $(i_\textup{c}=3)$, else,\\
        \indent $\mathbf{y}_i=[10101]$ if $c_{i+2}c_{i+1}c_i=\beta_1'\beta_3\gamma_1$  $(i_\textup{c}=4)$, else,\\
        \indent $\mathbf{y}_i=[01000]$ if $c_{i+2}c_{i+1}c_i=\beta_2'\beta_4\gamma_2$  $(i_\textup{c}=5)$, else,\\
        \indent $\mathbf{y}_i=[00100]$ if $c_{i+1}c_i=\beta_1'\alpha^2$  $(i_\textup{c}=6)$, else,\\
        \indent $\mathbf{y}_i=[01001]$ if $c_{i+1}c_i=\beta_1'\gamma_2$  $(i_\textup{c}=7)$, else,\\
        \indent $\mathbf{y}_i=[01110]$ if $c_{i+1}c_i=\beta_1'\alpha^5$  $(i_\textup{c}=8)$, else,\\
        \indent $\mathbf{y}_i=[10011]$ if $c_{i+1}c_i=\beta_1'\alpha^6$  $(i_\textup{c}=9)$, else,\\
        \indent $\mathbf{y}_i=[00101]$ if $c_{i+1}c_i=\beta_2'\alpha$  $(i_\textup{c}=10)$, else,\\
        \indent $\mathbf{y}_i=[01010]$ if $c_{i+1}c_i=\beta_2'\gamma_3$  $(i_\textup{c}=11)$, else,\\
        \indent $\mathbf{y}_i=[10011]$ if $c_{i+1}c_i=\beta_2'\gamma_1$  $(i_\textup{c}=12)$, else,\\
        \indent $\mathbf{y}_i=[01000]$ if $c_{i+1}c_i=\alpha\alpha$  $(i_\textup{c}=13)$, else,\\
        \indent $\mathbf{y}_i=[01101]$ if $c_{i+1}c_i=\alpha\alpha^2$  $(i_\textup{c}=14)$, else,\\
        \indent $\mathbf{y}_i=[10101]$ if $c_{i+1}c_i=\alpha\gamma_1$  $(i_\textup{c}=15)$, else,\\
        \indent $\mathbf{y}_i=[01000]$ if $c_{i+1}c_i=\alpha^4\gamma_2$  $(i_\textup{c}=16)$, else,\\
        \indent $\mathbf{y}_i=[00000]$ if $c_i \in \{0,1,\alpha\}$ $(i_\textup{c}=1)$, else, \\
        \indent $\mathbf{y}_i=[00101]$ if $c_i \in \{\alpha^2,\alpha^3,\alpha^4\}$ $(i_\textup{c}=1)$, else, \\
        \indent $\mathbf{y}_i=[01010]$ if $c_i \in \{\alpha^5,\alpha^6\}$ $(i_\textup{c}=1)$.\\
    The sets $\beta_1',\beta_2', \beta_3, \beta_4$ and $\gamma_1,\gamma_2,\gamma_3$ are defined as $\beta_1' \in \{0,1,\alpha^3\},\ \beta_2' \in \{\alpha^2, \alpha^5, \alpha^6\},\ \beta_3 \in \{\alpha^2,\alpha^5\},\  \beta_4 \in \{1,\alpha^3\}$ and  $\gamma_1 \in \{\alpha^5,\alpha^6\},\ \gamma_2 \in \{\alpha^3,\alpha^4\},\ \gamma_3 \in \{\alpha^2,\alpha^3\}$.
    
\begin{proof}
    Observe that all the equations of $g_{i,i_\textup{c}}(c_i)$, \eqref{eq:sc1_2}--\eqref{eq:sc16}, conform with the form given by \eqref{eq:Step_5_g_ic} for some nonnegative integers $\theta_{i,1}$ and $\theta_{i,2}$ such that $\theta_{i,1}\leq a_i$ and $ \theta_{i,2} \leq 3$. Therefore, we can write:
        \begin{equation}\label{eq:Step_5_g_ic}
            g_{i,i_\textup{c}}(c_i) =  \frac{a_i-\theta_{i,1}}{6}\left( N_8(i+1)-N_8(i) \right) +  \frac{\theta_{i,2}}{2} N_8(i).
        \end{equation}
    Since $\theta_{i,1}\leq a_i \leq \mathcal{L}(\alpha^6)=7$, we need only three merging variables $y_{i,1},\ y_{i,2},\ y_{i,3}$ to express $\theta_{i,1}$. From \eqref{eq:sc1_2}--\eqref{eq:sc16}:
        \begin{equation}\label{eq:theta_1}
            \theta_{i,1} = 4y_{i,1} +2y_{i,2}+ y_{i,3}.
        \end{equation}
     Similarly, since $ \theta_{i,2} \leq 3$, we use two more  merging variables $y_{i,4},\ y_{i,5}$ to express $\theta_{i,2}$. From \eqref{eq:sc1_2}--\eqref{eq:sc16}:
        \begin{equation}\label{eq:theta_2}
            \theta_{i,2} = 2y_{i,4}+ y_{i,5}.
        \end{equation}
    
    Noting that \eqref{eq:Step_5_g_ic} includes only the cardinalities $N_8(i+1)$ and $N_8(i)$, we utilize the merging functions $f_0^{\textup{mer}}(\cdot)$ and $f_1^{\textup{mer}}(\cdot)$ with the cardinalities $N_8(i+1)$ and $N_8(i)$, respectively. The unified expression representing the contribution of a symbol $c_i$ to the codeword index can be written as:
        \begin{equation}\label{eq:step_5_g_merg}
            g_{i}(c_i) = f_0^{\textup{mer}}(\cdot)N_8(i+1) + f_1^{\textup{mer}}(\cdot)N_8(i).
        \end{equation}
    Using \eqref{eq:theta_1} and \eqref{eq:theta_2}, \eqref{eq:Step_5_g_ic} can be written in the form of \eqref{eq:step_5_g_merg}, and the contribution of a symbol $c_i$ to the codeword index $g(\mathbf{c})$ in Theorem~\ref{thr:step_5} is obtained:
        \begin{equation}\label{eq:step_5_g_i}
           g_{i}(c_i) = \left( \frac{a_i-\theta_{i,1}}{6}\right) N_8(i+1) + \left( \frac{\theta_{i,1} +3\theta_{i,2} -a_i}{6}\right) N_8(i)
        \end{equation}
    Finally, observe that \eqref{eq:step_5_g_i} is consistent with all cases of $g_{i,i_\textup{c}}(c_i)$ given by \eqref{eq:sc1_2}--\eqref{eq:sc16} once the correct merging variable vector $\mathbf{y}_i$ is selected following the rules in Theorem~\ref{thr:step_5}.
\end{proof}
\end{theorem}

\section{Simulation Results and Reconfigurability}\label{sec_sims}

In this section, we introduce the finite-length rates of OT-LOCO codes. Then, we discuss the necessary modifications on the TDMR model in Section~\ref{sec_motiv} to perform simulations of the coded system. We introduce and discuss performance plots that demonstrate the gains achieved by OT-LOCO codes, and we also show plots describing the process of reconfigurability between OP-LOCO and OT-LOCO codes in a TDMR system.

Before discussing finite-length rates, we need to first address \textit{bridging} and \textit{self-clocking} in the proposed codes.

\textbf{OT-LOCO bridging:} In fixed-length constrained codes, bridging is required to prevent forbidden patterns from appearing at the transition between consecutive codewords in a stream \cite{ahh_loco, ahh_general}. Bridging in any LOCO code does not affect the asymptotic rate since the number of bridging symbols does not grow with the code length $m$. However, care should be taken while choosing or devising a bridging scheme since it affects finite-length rates.

Recall the set of forbidden patterns:
\begin{align}
    \mathcal{OT}^8 \triangleq \{
    & \alpha\beta_1,\ \beta_1\alpha,\ \alpha^4\beta_2,\ \beta_2\alpha^4,\ \bar\beta_1'\beta_3\beta_1',\   \bar\beta_2'\beta_4\beta_2',
    \ \forall \beta_1 \in \{0,1,\alpha^3,\alpha^4\},\ \forall \bar \beta_1', \beta_1' \in \{0,1,\alpha^3\}, \nonumber \\
    &\ \forall \beta_2 \in \{\alpha,\alpha^2,\alpha^5,\alpha^6\},\ \forall \bar \beta_2', \beta_2' \in \{\alpha^2,\alpha^5,\alpha^6\},
    \ \forall \beta_3 \in \{\alpha^2,\alpha^5\},\ \forall \beta_4 \in \{1,\alpha^3\}
    \}. \nonumber
\end{align}
For the sake of brevity, we here use the following format when studying the bridging between two consecutive OT-LOCO codewords in $\mathcal{OTC}^8_{m}$ ``right-most symbols of codeword at $t$ -- bridging pattern -- left-most symbols of codeword at $t+1$''. Denote the bridging sequence by $\mathbf{d}$, whose length is unspecified yet. Note that the length of $\mathbf{d}$ has to be fixed for all possible cases since our coding scheme is of fixed-length.

Consider the following case:
\begin{equation}
\beta_1'\beta_3 - \mathbf{d} - \beta_4\beta_2'.
\end{equation}
We first examine the situation of $\mathbf{d}$ of length $1$. Because of the right-most symbols of the codeword at $t$, $\mathbf{d}$ cannot be in $\{0, 1, \alpha^3, \alpha^4\}$. Moreover, because of the left-most symbols of the codeword at $t+1$, $\mathbf{d}$ cannot be in $\{\alpha, \alpha^2, \alpha^5, \alpha^6\}$. Therefore, $\mathbf{d}$ cannot be of length $1$. Next, we examine the situation of $\mathbf{d}$ of length $2$, i.e., $\mathbf{d} = d_1 d_0$. Because of the right-most symbols of the codeword at $t$, $d_1$ must be in $\{\alpha, \alpha^2, \alpha^5, \alpha^6\}$. Moreover, because of the left-most symbols of the codeword at $t+1$, $d_0$ must be in $\{0, 1, \alpha^3, \alpha^4\}$. To be able to concatenate $d_1$ and $d_0$ without creating a $2$-tuple forbidden pattern, $d_1 \neq \alpha$ and $d_0 \neq \alpha^4$. Therefore, 
\begin{align}\label{eqn_9ford_1}
&\mathbf{d} = d_1d_0 = \bar\beta_2'\bar\beta_1', \nonumber \\
&\bar\beta_2' \in \{\alpha^2, \alpha^5, \alpha^6\} \textup{ and } \bar\beta_1' \in \{0, 1, \alpha^3\}.
\end{align}
It can be shown in a similar way that for the case:
\begin{equation}
\beta_2'\beta_4 - \mathbf{d} - \beta_3\beta_1',
\end{equation}
we will have:
\begin{align}\label{eqn_9forb_2}
&\mathbf{d} = d_1d_0 = \bar\beta_1'\bar\beta_2', \nonumber \\
&\bar\beta_1' \in \{0, 1, \alpha^3\} \textup{ and } \bar\beta_2' \in \{\alpha^2, \alpha^5, \alpha^6\}.
\end{align}

By checking all the possible cases for the right-most symbols of the codeword at $t$ and the left-most symbols of the codeword at $t+1$, one can conclude that the aforementioned two cases result in the highest level of restrictions on $\mathbf{d}$; that is, $\mathbf{d}$ has to be of length $2$ and there are only $3 \times 3 = 9$ possible options for it as shown in \eqref{eqn_9ford_1} and \eqref{eqn_9forb_2}. Because OT-LOCO codes are fixed-length codes, we have to always abide by these restrictions. Thus, we specify all bridging patterns to use for all cases based on the rules:
\begin{enumerate}
\item Each case has $9$ possible bridging patterns out of which, we will use only $8$.
\item Bridging patterns must prevent forbidden patterns from appearing at the transition between codewords.
\item Each bridging pattern $\mathbf{d}$ is of length $2$, i.e., $\mathbf{d} = d_1d_0$.
\item Each of $d_1$ and $d_0$ is either $\bar\beta_1'$ or $\bar\beta_2'$.
\end{enumerate}

\begin{table}
\caption{All Bridging Scenarios of OT-LOCO Codes. Scenarios Are Separated by Two Horizontal Lines. Scenario Priority Reduces From Top to Bottom. Symbol $x_j$ Refers to ``Does Not Matter''.}
\vspace{-0.5em}
\centering
\renewcommand{\arraystretch}{1.5}
\scalebox{1.00}
{
\begin{tabular}{|c|c|c|}
\hline
\multirow{4}{*}{Scenario~1} & $x_1\alpha - \bar\beta_2' \bar\beta_1' - \beta_1x_2$ & $x_1\beta_1 - \bar\beta_1' \bar\beta_2' - \alpha x_2$ \\
\cline{2-2}\cline{3-3}
 & $x_1\alpha^4 - \bar\beta_1' \bar\beta_2' - \beta_2x_2$ & $x_1\beta_2 - \bar\beta_2' \bar\beta_1' - \alpha^4x_2$ \\
\cline{2-2}\cline{3-3}
 & $\beta_1'\beta_3 - \bar\beta_2' \bar\beta_1' - \beta_1'x_1$ & $x_1\beta_1' - \bar\beta_1' \bar\beta_2' - \beta_3\beta_1'$ \\
\cline{2-2}\cline{3-3}
 & $\beta_2'\beta_4 - \bar\beta_1' \bar\beta_2' - \beta_2'x_1$ & $x_1\beta_2' - \bar\beta_2' \bar\beta_1' - \beta_4\beta_2'$ \\
\hline
\hline
\multirow{4}{*}{\makecell{Else \\ Scenario~2}} & $x_1\alpha - \bar\beta_2' \bar\beta_2' - x_2x_3$ & $x_1x_2 - \bar\beta_2' \bar\beta_2' - \alpha x_3$ \\
\cline{2-2}\cline{3-3}
 & $x_1\alpha^4 - \bar\beta_1' \bar\beta_1' - x_2x_3$ & $x_1x_2 - \bar\beta_1' \bar\beta_1' - \alpha^4x_3$ \\
\cline{2-2}\cline{3-3}
 & $\beta_1'\beta_3 - \bar\beta_2' \bar\beta_2' - x_1x_2$ & $x_1x_2 - \bar\beta_2' \bar\beta_2' - \beta_3\beta_1'$ \\
\cline{2-2}\cline{3-3}
 & $\beta_2'\beta_4 - \bar\beta_1' \bar\beta_1' - x_1x_2$ & $x_1x_2 - \bar\beta_1' \bar\beta_1' - \beta_4\beta_2'$ \\
\hline
\hline
\multirow{2}{*}{\makecell{Else \\ Scenario~3}} & $x_1\beta_1' - \bar\beta_2' \bar\beta_2' - \beta_1'x_2$ & $x_1\beta_2' - \bar\beta_1' \bar\beta_1' - \beta_2'x_2$ \\
\cline{2-2}\cline{3-3}
 & $x_1\beta_1' - \bar\beta_1' \bar\beta_2' - \beta_2'x_2$ & $x_1\beta_2' - \bar\beta_2' \bar\beta_1' - \beta_1'x_2$ \\
\hline
\hline
{\makecell{Otherwise \\ Scenario~4}} & \multicolumn{2}{|c|}{$x_1x_2 - \bar\beta_1' \bar\beta_2' - x_3x_4 $} \\
\hline
\end{tabular}}
\label{table_2}
\vspace{-0.5em}
\end{table}

Table~\ref{table_2} details all bridging scenarios of OT-LOCO codes. Furthermore, there is an additional idea we apply in OT-LOCO bridging, which is to use bridging symbols to encode binary input message bits in order to increase the finite-length rate (see also \cite{ahh_rr}). In particular, since we have $9$ possible bridging patterns for every case, we can encode up to $3$ additional input message bits via these bridging symbols. We encode $3$ additional bits in our setup, and that is why we only save and use $2^3 = 8$ bridging patterns for each case within the above four scenarios. It is important to highlight that the impact of such a simple idea on the finite-length rate is remarkable.

\textbf{OT-LOCO self-clocking:} Self-clocking is needed in magnetic recording systems in order that the read head keeps track of grain boundaries (cross track boundaries in TDMR), which allows the system to have self-calibration \cite{siegel_mr, ahh_loco}.

In order to achieve self-clocking, very long same-symbol sequences should not be allowed. Typically, we eliminate some same-symbol codewords, e.g., $\mathbf{0}^m$ and $\boldsymbol{\alpha}^m$ (all $0$ and all $\alpha$ codewords), from $\mathcal{OTC}^8_{m}$ to achieve self-clocking \cite{tang_bahl, ahh_loco}. However, we do not need to do that here. The reason is that our bridging in Table~\ref{table_2} is devised such that a transition from a symbol to a different symbol occurs within the bridging interval or/and directly before/after the bridging interval. This makes our OT-LOCO codes \textit{intrinsically self-clocked}. Denote the maximum number of consecutive GF$(8)$ symbols ($3$-bit columns) that are identical in an OT-LOCO stream coded via $\mathcal{OTC}^8_{m}$ and stored in a TDMR device by $k_\textup{eff}^\textup{ot}$. Then,
\begin{equation}
k_\textup{eff}^\textup{ot} = m+4.
\end{equation}

\textbf{OT-LOCO rates:} The rate of an OT-LOCO code $\mathcal{OTC}^8_{m}$ is the number of message bits divided by the number of coded symbols. We have two types of messages bits; the ones converted into an OT-LOCO codeword and the ones encoded within bridging. The number of bits converted into an OT-LOCO codeword is given by:
\begin{equation}\label{eqn_s}
s = \left \lfloor \log_2 (N_8(m)) \right \rfloor,
\end{equation}
while we encode $3$ bits for each bridging pattern as illustrated above. Observe that $s$ is also the size of the adder that executes the encoding-decoding rule, and therefore, it dictates the complexity of encoding-decoding. We also have two types of coded symbols; $m$ symbols for an OT-LOCO codeword and two more symbols for the following bridging pattern. Combining that with \eqref{eqn_s} gives the rate of an OT-LOCO code $\mathcal{OTC}^8_{m}$ as follows:
\begin{equation}\label{eqn_rate_ot}
R_\textup{OT-LOCO} = \frac{s+3}{m+2} = \frac{\left \lfloor \log_2 (N_8(m)) \right \rfloor +3}{m+2}.
\end{equation}
The normalized rate of an OT-LOCO code $\mathcal{OTC}^8_{m}$ is then:
\begin{equation}\label{eqn_raten_ot}
R_\textup{OT-LOCO}^\textup{n} = \frac{s+3}{3(m+2)} = \frac{\left \lfloor \log_2 (N_8(m)) \right \rfloor +3}{3(m+2)}.
\end{equation}

\begin{table}
\caption{Rates, Normalized Rates, and Adder Sizes of OT-LOCO codes $\mathcal{OTC}^8_m$ for Different Values of $m$. The Capacity Is $2.5494$, and the Normalized Capacity Is $0.8498$.}
\vspace{-0.5em}
\centering
\scalebox{1.00}
{
\begin{tabular}{|c|c|c|c|}
\hline
\makecell{$m$} & \makecell{$R_{\textup{OT-LOCO}}$} & \makecell{$R_{\textup{OT-LOCO}}^{\textup{n}}$}  & \makecell{Adder size} \\
\hline
$10$ & $2.4167$ & $0.8056$ & $26$ bits \\
\hline
$14$ & $2.4375$ & $0.8125$ & $36$ bits \\
\hline
$21$ & $2.4783$ & $0.8261$ & $54$ bits \\
\hline
$30$ & $2.5000$ & $0.8333$ & $77$ bits \\
\hline
$50$ & $2.5192$ & $0.8397$ & $128$ bits \\
\hline
$81$ & $2.5301$ & $0.8434$ & $207$ bits \\
\hline
\end{tabular}}
\label{table_3}
\vspace{-0.5em}
\end{table}

Observe that all codewords satisfying the constraint are included in $\mathcal{OTC}^8_m$. Moreover, the number of bridging symbols does not grow with $m$. Therefore,
\begin{equation}
\lim_{m \to \infty} R_\textup{OT-LOCO} = \lim_{m \to \infty} \frac{\left \lfloor \log_2 (N_8(m)) \right \rfloor +3}{m+2} = C = 2.5494.
\end{equation}
This means OT-LOCO codes are \textit{capacity-achieving} constrained codes. Table~\ref{table_3} gives the rates, normalized rates, and adder sizes of OT-LOCO codes $\mathcal{OTC}^8_m$ for different values of $m$. Observe how finite-length rates approach capacity.

Bridging, self-clocking, and finite-length rates are all part of the \textit{sixth step} of the general method in \cite{ahh_general}. We skip the remainder of the sixth step of the general method and refer the reader to our previous work to see how the encoding-decoding algorithms are derived from the encoding-decoding rule. Having said that, we have implemented, tested, and used such OT-LOCO encoding-decoding algorithms as discussed below.\vspace{+0.7em}

Next, we discuss our coded TDMR system setup. We have the writing setup, the channel setup, and the reading setup. For brevity, we omit some details that are already clarified for the uncoded TDMR system setup in Section~\ref{sec_motiv}.

\textbf{Writing setup:} We generate random binary input messages. Then, we encode each message $\mathbf{b}$ of length $s$ into the corresponding $8$-ary OT-LOCO codeword $\mathbf{c}$ of length $m$. After each message, $3$ free input message bits are used to specify $2$ bridging symbols $\mathbf{d}$. Each GF$(8)$ symbol in the OT-LOCO codeword is converted into a $3 \times 1$ column of binary bits according to the mapping-demapping in \eqref{eqn_gf8map}. Consequently, a codeword of length $m$ will be written over a grid of size $3 \times m$. Two bridging columns separate each two consecutive OT-LOCO codewords. Before writing, level-based signaling is applied, which converts each $0$ into $-1$, each $1$ into $+1$. Upon writing, these $-1$ and $+1$ values will be updated to values depending on $TW$ and $BP$.

We use the following OT-LOCO codes in the simulations:
\begin{itemize}
\item The code $\mathcal{OTC}^8_{23}$ with codeword length $m = 23$, message length $s = 59$ plus $3$ free input bits, and normalized rate $R^{\textup{n}}_{\textup{OT-LOCO}} = 0.8267$.
\item The code $\mathcal{OTC}^8_{14}$ with codeword length $m = 14$, message length $s = 36$ plus $3$ free input bits, and normalized rate $R^{\textup{n}}_{\textup{OT-LOCO}} = 0.8125$.
\end{itemize}
We also use the following OP-LOCO codes in the simulations (see \cite{ahh_general} for more details):
\begin{itemize}
\item The code $\mathcal{OPC}^8_{23}$ with codeword length $m = 23$, message length $s = 67$, and normalized rate $R^{\textup{n}}_{\textup{OP-LOCO}} = 0.9306$.
\item The code $\mathcal{OPC}^8_{14}$ with codeword length $m = 14$, message length $s = 40$, and normalized rate $R^{\textup{n}}_{\textup{OP-LOCO}} = 0.8889$.
\end{itemize}

It is important to perform fair comparisons between the OT-LOCO coded and the uncoded settings, and this will not happen if we use in the coded setting the same $TW$ and $BP$ of the uncoded setting. In order to fix the energy, i.e., keep the energy per input message bit in the coded setting the same as it is in the uncoded setting, we obtain $TW$ and $BP$ of the coded setting via scaling both $TW$ and $BP$ of the uncoded setting by $\sqrt{R^{\textup{n}}_{\textup{OT-LOCO}}}$, respectively. The same is also done for the OP-LOCO coded setting. This scaling is skipped in the reconfigurability plots since the system switches between two coded settings. In particular, we switch from OP-LOCO to OT-LOCO coding as the TDMR device gets older.

\textbf{Channel setup:} Our baseline channel model is the TDMR model in \cite{mohsen_tdmr}, which is a Voronoi model. Here, we only consider media noise/interference. We modify this model such that it is suitable for a wide read head that reads data from $3$ adjacent down tracks simultaneously. In particular, in each group of $3$ adjacent down tracks, the upper and lower tracks in our model have additional protection from interference in the cross track direction. Thus, the middle down track in each group suffers from the highest level of interference \cite{chan_tdmr, bd_tdmr}.

In the simulations, we have two sweep setups. First, we sweep the TD channel density $D_{\textup{TD}}$ given in \eqref{eqn_tddensity}. The details of the sweep are illustrated in Section~\ref{sec_motiv}. Second, we sweep the TD bit energy metric $E_{\textup{TD}} = TW \times BP$. The sweep is done such that the TD density $D_{\textup{TD}}$ is fixed at $1.00$, and the details of the sweep are also illustrated in Section~\ref{sec_motiv}.

The input to the channel is $3 \times m$ grids of coded bits along with their $2$ bridging columns after signaling is applied.\footnote{The word ``coded'' in this section means ``OT-LOCO coded'' unless otherwise explicitly stated (OP-LOCO for reconfigurability).} The output from the channel is these $3 \times m$ grids along with the bridging columns after Voronoi media noise/interference is applied, taking into account the aforementioned protection of the upper and lower tracks in each group of $3$ down tracks. Mathematically, the channel effect is equivalent to applying the TD convolution between the $3 \times m$ input grids with their bridging columns and the $3 \times 3$ read-head impulse response with media noise.

\textbf{Reading setup:} Outputs of the channel, which are $3 \times m$ grids, are read based on hard decision applied to the value of each entry. If the value of the entry is less than or equal to zero, the corresponding bit is read as $0$. In contrast, if the value of the entry is greater than zero, the corresponding bit is read as $1$. The same applies to the two bridging columns. We then convert coded columns of $3$ bits each into GF$(8)$ symbols according to \eqref{eqn_gf8map}. Each coded $8$-ary sequence of length $m$ is then checked for constraint violation. Whenever the constraint is violated, a frame error is counted. Otherwise, the OT-LOCO codeword $\widehat{\bold{c}}$ passes through the decoding algorithm to obtain the corresponding binary message $\widehat{\bold{b}}$. If $g(\widehat{\bold{c}}) \geq 2^s$, a frame error is counted. If $\widehat{\bold{b}} \neq \bold{b}$, which is implied by $\widehat{\mathbf{c}} \neq \mathbf{c}$, a frame error is counted. Each two $8$-ary bridging symbols are decoded into $3$ binary bits according to our bridging rules. If these bits do not match the original ones, a frame error is also counted.

We have frame error rate (FER) and bit error rate (BER) plots. Bit errors, as the name tells, are just counted when~the hard-decision value of the channel output does not match the input value to the channel ($+1$ or $-1$). Observe that the additional protection on the upper and lower tracks makes the average amount of interference required to cause an error on these tracks higher than that of the middle track. That is why in our plots, we show the performance on all $3$ tracks as well as the performance on the middle track only.\vspace{+0.7em}

\begin{figure}
\vspace{-0.4em}
\center
\includegraphics[trim={0.0in 0.0in 0.0in 0.0in}, width=3.2in]{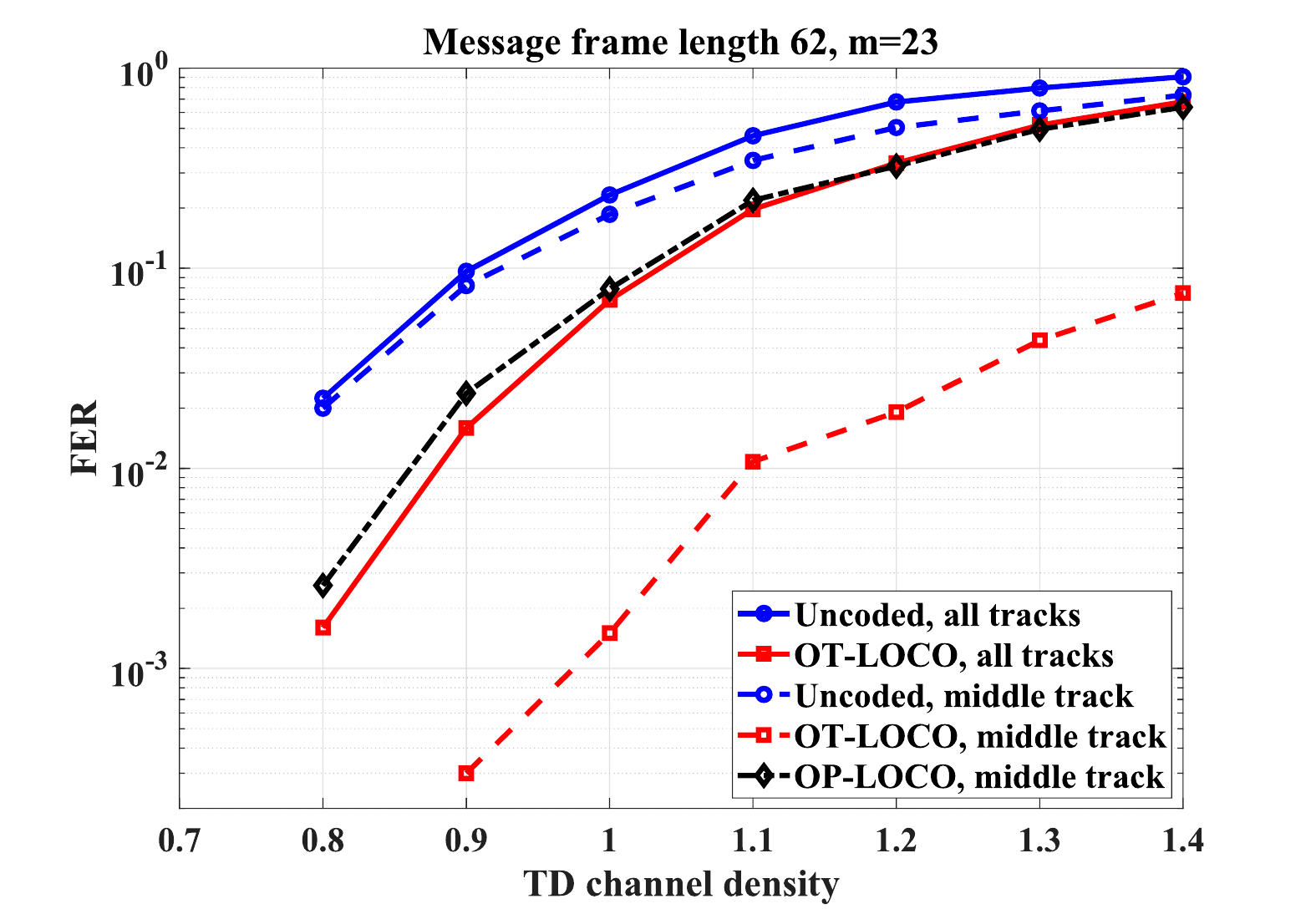}
\vspace{-0.7em}
\caption{FER versus TD density comparisons between $\mathcal{OTC}^8_{23}$-coded setting and uncoded setting for all tracks, between $\mathcal{OTC}^8_{23}$-coded setting and uncoded setting for middle track, and between $\mathcal{OTC}^8_{23}$-coded setting and $\mathcal{OPC}^8_{23}$-coded setting for middle track.}
\label{fig_perf1}
\vspace{-0.3em}
\end{figure}

\begin{figure}
\vspace{-0.2em}
\center
\includegraphics[trim={0.0in 0.0in 0.0in 0.0in}, width=3.2in]{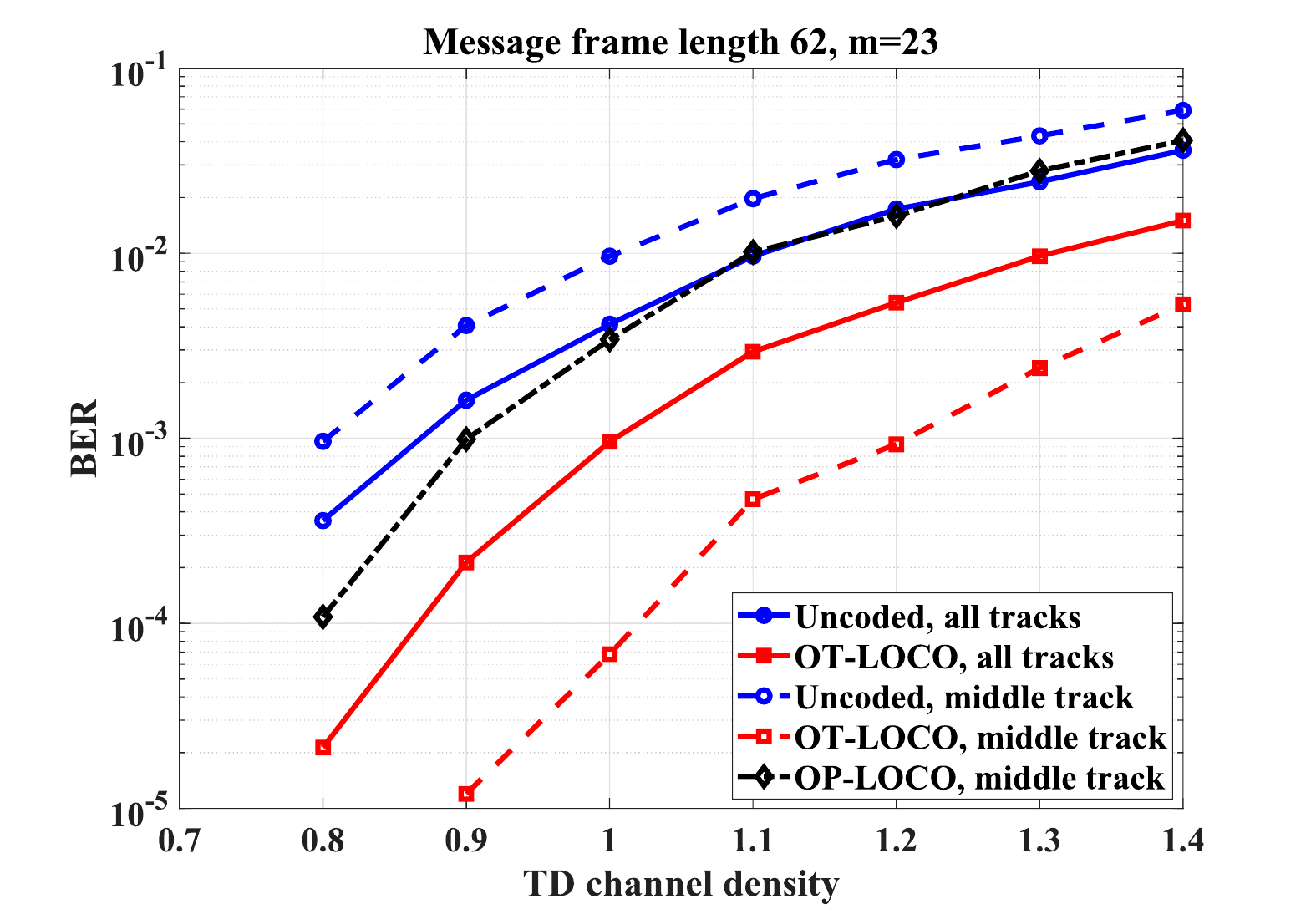}
\vspace{-0.7em}
\caption{BER versus TD density comparisons between $\mathcal{OTC}^8_{23}$-coded setting and uncoded setting for all tracks, between $\mathcal{OTC}^8_{23}$-coded setting and uncoded setting for middle track, and between $\mathcal{OTC}^8_{23}$-coded setting and $\mathcal{OPC}^8_{23}$-coded setting for middle track.}
\label{fig_perf2}
\vspace{-0.4em}
\end{figure}

Fig.~\ref{fig_perf1} and Fig.~\ref{fig_perf2} demonstrate the gains achieved by OT-LOCO codes compared with the uncoded setting in TDMR upon sweeping the TD channel density $D_{\textup{TD}}$. In particular, the figures compare the system coded via the OT-LOCO code $\mathcal{OTC}^8_{23}$ with the uncoded system at fixed energy per input message bit. Fig.~\ref{fig_perf1} introduces the FER performance for all tracks and for the middle track, while Fig.~\ref{fig_perf2} introduces the BER performance for all tracks and for the middle track.

We first discuss Fig.~\ref{fig_perf1}. At FER $\approx 2.0 \times 10^{-2}$, the OT-LOCO code achieves a TD density gain of about $15\%$ for all tracks. At the same FER, the OT-LOCO code achieves a TD density gain of about $50\%$ for the middle track. At $D_{\textup{TD}}=0.8$, the OT-LOCO code achieves an FER performance gain of about $1.15$ orders of magnitude for all tracks. At $D_{\textup{TD}}=0.9$, the OT-LOCO code achieves an FER performance gain of about $2.44$ orders of magnitude for the middle track. Next, we discuss Fig.~\ref{fig_perf2}. At BER $\approx 3.6 \times 10^{-4}$, the OT-LOCO code achieves a TD density gain of about $17\%$ for all tracks. At BER $\approx 1.0 \times 10^{-3}$, the OT-LOCO code achieves a TD density gain of about $50\%$ for the middle track. At $D_{\textup{TD}}=0.8$, the OT-LOCO code achieves a BER performance gain of about $1.23$ orders of magnitude for all tracks. At $D_{\textup{TD}}=0.9$, the OT-LOCO code achieves a BER performance gain of about $2.53$ orders of magnitude for the middle track.

A major observation from Fig.~\ref{fig_perf1} and Fig.~\ref{fig_perf2} is that we could not collect any frame errors nor any bit errors at TD densities below $0.8$ out of $10{,}000$ frames simulated. This demonstrates the \textbf{elimination of media noise and interference effects} at  practical TD densities in the TDMR system via OT-LOCO codes with the acceptable rate of $0.8267$. Another pivotal observation from Fig.~\ref{fig_perf1} and Fig.~\ref{fig_perf2} is that OT-LOCO codes remarkably outperform OP-LOCO codes at the same density and the same setup. At $D_{\textup{TD}}=0.9$, the OT-LOCO code achieves an FER performance gain of about $1.90$ orders of magnitude for the middle track compared with the OP-LOCO code. At the same density, the OT-LOCO code achieves a BER performance gain of about $1.92$ orders of magnitude for the middle track compared with the OP-LOCO code.

\begin{figure}
\vspace{-0.4em}
\center
\includegraphics[trim={0.0in 0.0in 0.0in 0.0in}, width=3.2in]{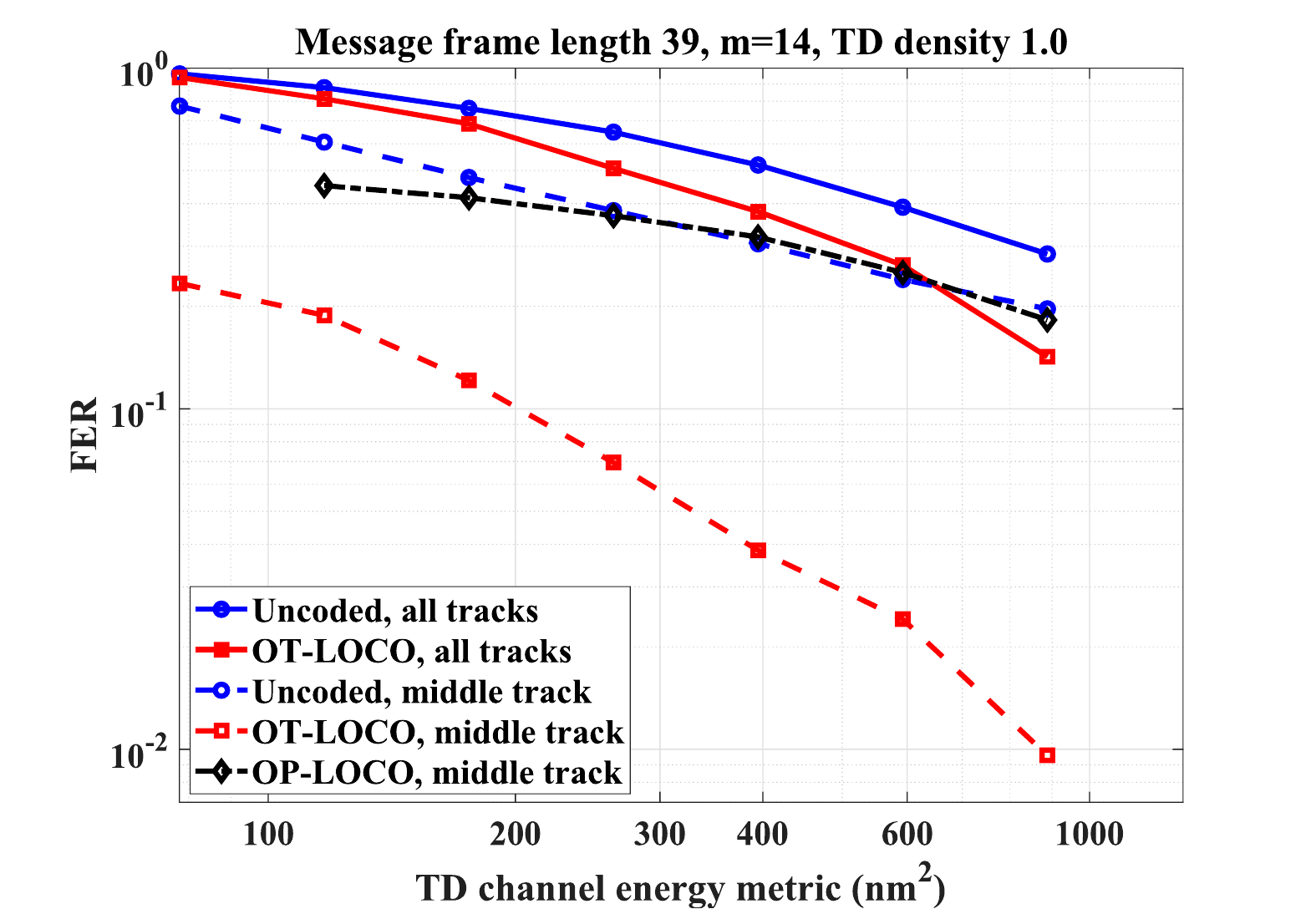}
\vspace{-0.7em}
\caption{FER versus TD energy metric comparisons between $\mathcal{OTC}^8_{14}$-coded setting and uncoded setting for all tracks, between $\mathcal{OTC}^8_{14}$-coded setting and uncoded setting for middle track, and between $\mathcal{OTC}^8_{14}$-coded setting and $\mathcal{OPC}^8_{14}$-coded setting for middle track.}
\label{fig_perf3}
\vspace{-0.3em}
\end{figure}

\begin{figure}
\vspace{-0.3em}
\center
\includegraphics[trim={0.0in 0.0in 0.0in 0.0in}, width=3.2in]{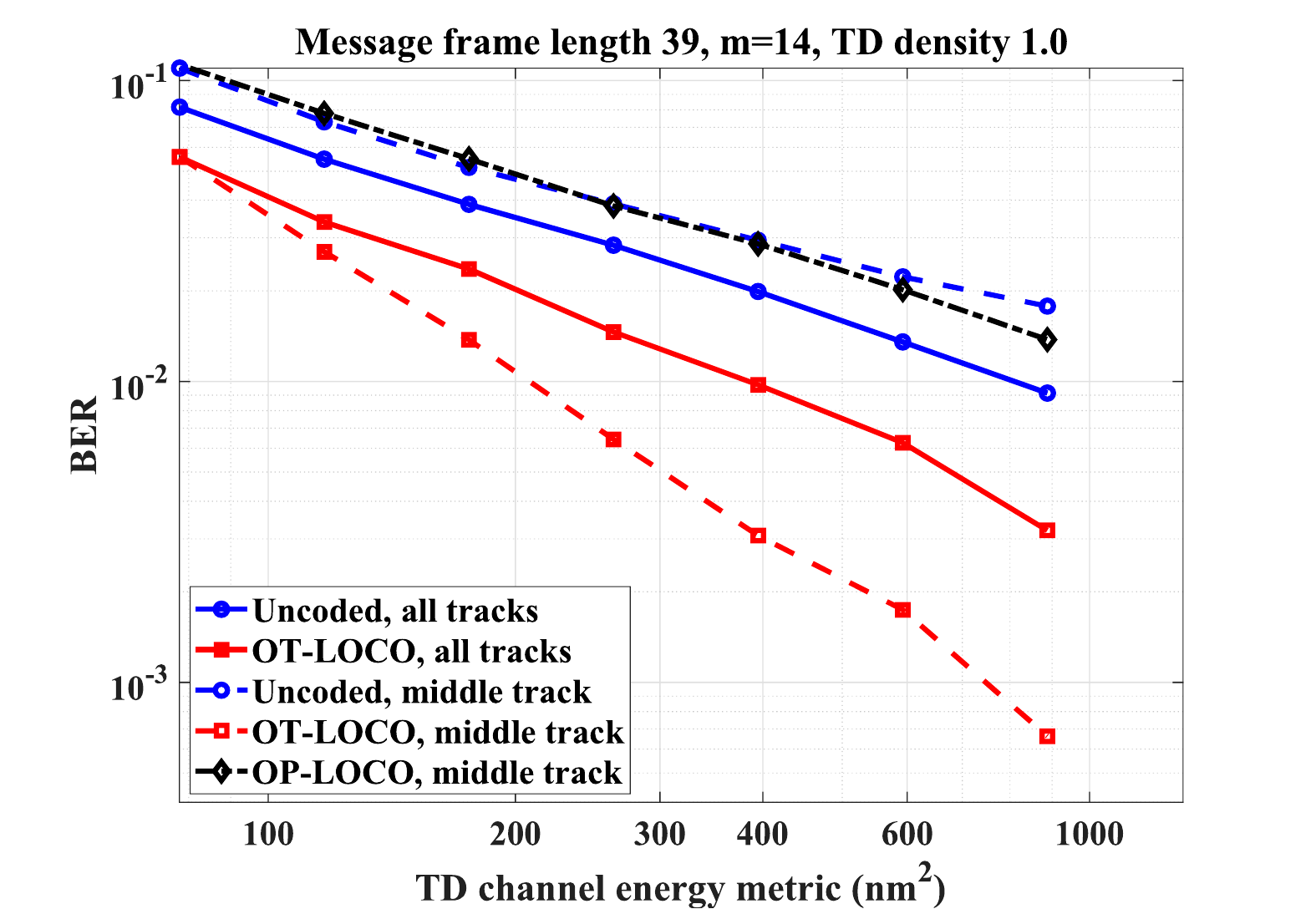}
\vspace{-0.7em}
\caption{BER versus TD energy metric comparisons between $\mathcal{OTC}^8_{14}$-coded setting and uncoded setting for all tracks, between $\mathcal{OTC}^8_{14}$-coded setting and uncoded setting for middle track, and between $\mathcal{OTC}^8_{14}$-coded setting and $\mathcal{OPC}^8_{14}$-coded setting for middle track.}
\label{fig_perf4}
\vspace{-0.4em}
\end{figure}

Fig.~\ref{fig_perf3} and Fig.~\ref{fig_perf4} demonstrate the gains achieved by OT-LOCO codes compared with the uncoded setting in TDMR upon sweeping the TD channel energy metric $E_{\textup{TD}}$. In particular, the figures compare the system coded via the OT-LOCO code $\mathcal{OTC}^8_{14}$ with the uncoded system at fixed energy per input message bit and at $D_{\textup{TD}}=1.0$. Fig.~\ref{fig_perf3} introduces the FER performance for all tracks and for the middle track, while Fig.~\ref{fig_perf4} introduces the BER performance for all tracks and for the middle track.

We first discuss Fig.~\ref{fig_perf3}. At FER $\approx 2.8 \times 10^{-1}$, the OT-LOCO code achieves a TD energy gain of about $38\%$ for all tracks. At FER $\approx 2.0 \times 10^{-1}$, the OT-LOCO code achieves a TD energy gain of about $87\%$ for the middle track. At $E_{\textup{TD}}=888.6$, the OT-LOCO code achieves an FER performance gain of about $0.30$ of an order of magnitude for all tracks. At the same $E_{\textup{TD}}$, the OT-LOCO code achieves an FER performance gain of about $1.30$ orders of magnitude for the middle track. Next, we discuss Fig.~\ref{fig_perf4}. At BER $\approx 9.2 \times 10^{-3}$, the OT-LOCO code achieves a TD energy gain of about $52\%$ for all tracks. At BER $\approx 1.8 \times 10^{-2}$, the OT-LOCO code achieves a TD energy gain of about $83\%$ for the middle track. At $E_{\textup{TD}}=888.6$, the OT-LOCO code achieves a BER performance gain of about $0.46$ of an order of magnitude for all tracks. At the same $E_{\textup{TD}}$, the OT-LOCO code achieves a BER performance gain of about $1.43$ orders of magnitude for the middle track.

Another pivotal observation from Fig.~\ref{fig_perf3} and Fig.~\ref{fig_perf4} is that OT-LOCO codes remarkably outperform OP-LOCO codes at the same energy and the same setup. At $E_{\textup{TD}}=888.6$, the OT-LOCO code achieves an FER performance gain of about $1.28$ orders of magnitude for the middle track compared with the OP-LOCO code. At the same energy metric, the OT-LOCO code achieves a BER performance gain of about $1.32$ orders of magnitude for the middle track compared with the OP-LOCO code. An intriguing observation is that OP-LOCO codes achieve almost no FER/BER gain compared with the uncoded setting in the low energy regime, unlike OT-LOCO codes.

\begin{remark}
Note that Fig.~\ref{fig_perf1} and Fig.~\ref{fig_perf3} show that the uncoded FER performance for the middle track is a little better than that for all tracks. The reason is that a frame error is counted if a bit error occurs on any of the $3$ down tracks. On the contrary, Fig.~\ref{fig_perf2} and Fig.~\ref{fig_perf4} show that the uncoded BER performance for the middle track is worse than that for all tracks. The reason is that media noise and interference affect the middle track the most compared with the upper and lower tracks.
\end{remark}

\begin{figure}
\vspace{-0.4em}
\center
\includegraphics[trim={0.0in 0.0in 0.0in 0.0in}, width=3.2in]{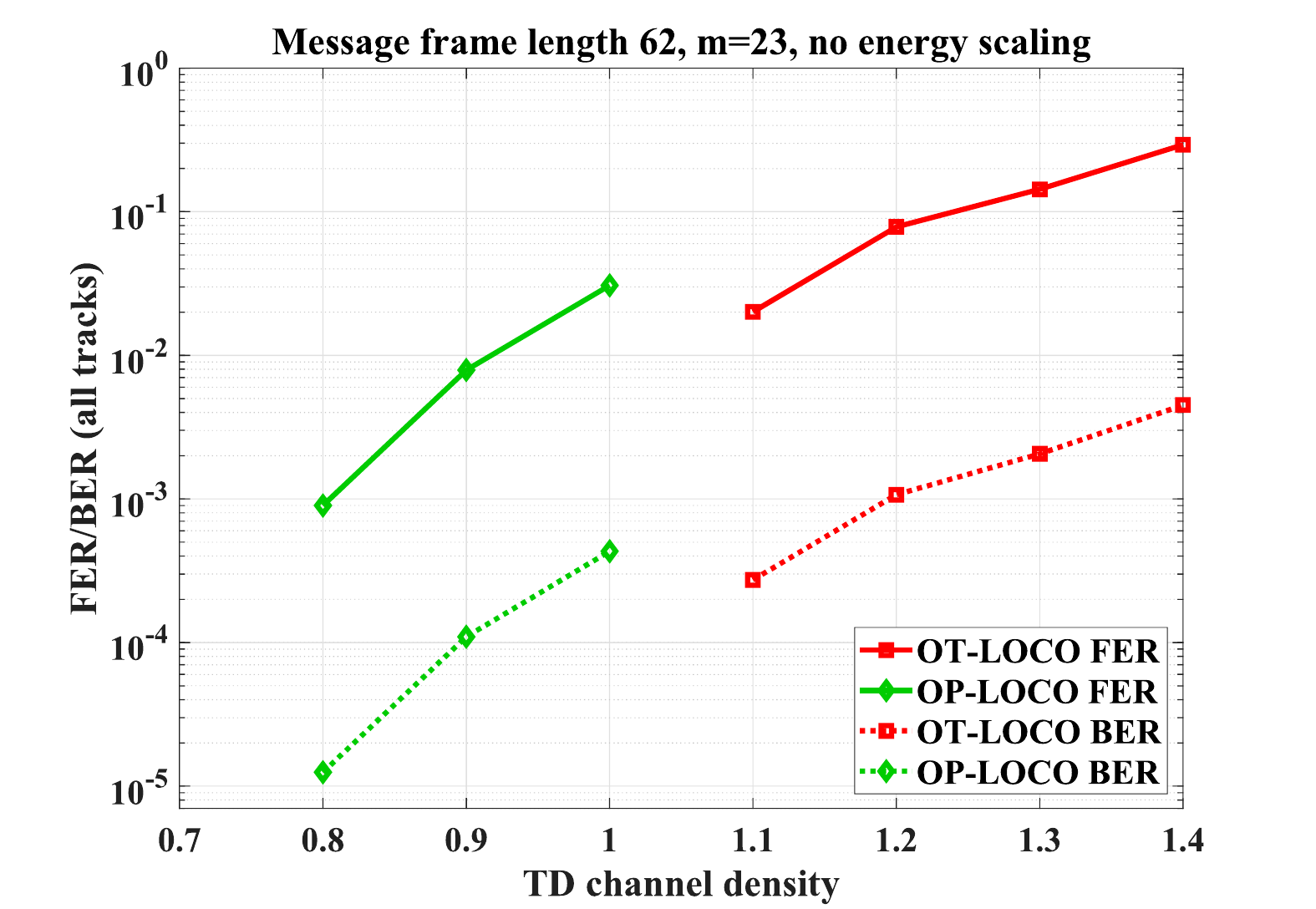}
\vspace{-0.7em}
\caption{FER/BER reconfiguration plots using $\mathcal{OPC}^8_{23}$ for low density and $\mathcal{OTC}^8_{23}$ for high density.}
\label{fig_reconf1}
\vspace{-0.3em}
\end{figure}

\begin{figure}
\vspace{-0.5em}
\center
\includegraphics[trim={0.0in 0.0in 0.0in 0.0in}, width=3.2in]{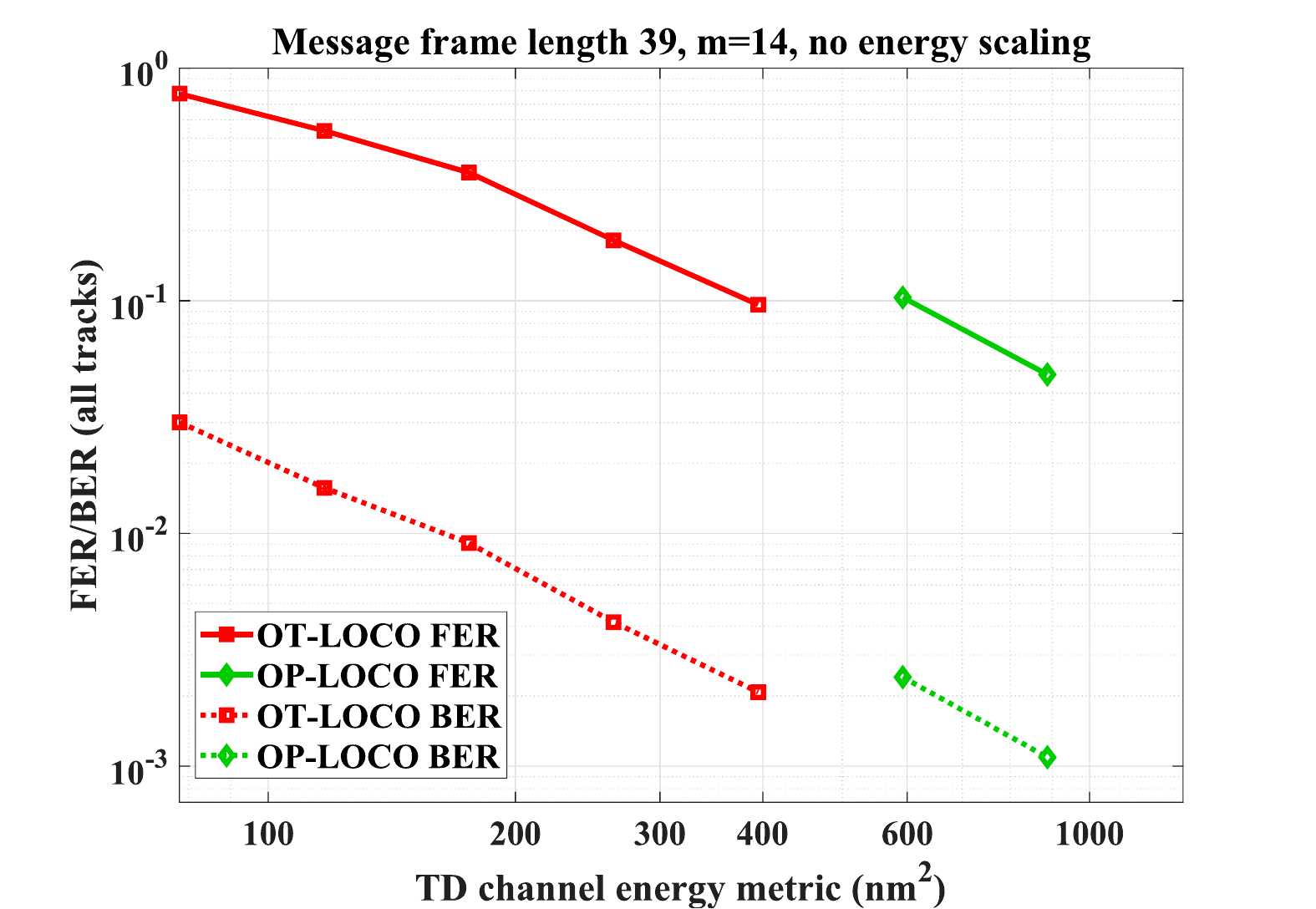}
\vspace{-0.7em}
\caption{FER/BER reconfiguration plots using $\mathcal{OPC}^8_{14}$ for high energy and $\mathcal{OTC}^8_{14}$ for low energy.}
\label{fig_reconf2}
\vspace{-0.4em}
\end{figure}

Fig.~\ref{fig_reconf1} and Fig.~\ref{fig_reconf2} demonstrate the concept of reconfigurability in the TDMR system. The idea is that reconfiguring LOCO codes is as easy as reprogramming an adder since their encoding and decoding follow a simple rule that links the codeword to its index (for OT-LOCO codes, it is \eqref{eqn_ruleot}). In order to further preserve storage capacity, we can use an OP-LOCO code when the device is relatively fresh, then switch to an OT-LOCO code as the device gets older. Note that OP-LOCO codes require less redundancy, and note also that PIS patterns dominate the error profile of the TDMR device at low density and/or high energy as shown in Section~\ref{sec_motiv}.

We sweep the TD channel density $D_{\textup{TD}}$ in Fig.~\ref{fig_reconf1}. The two codes we use are the OP-LOCO code $\mathcal{OPC}^8_{23}$ of rate $0.9306$ and the OT-LOCO code $\mathcal{OTC}^8_{23}$ of rate $0.8267$. The reconfiguration criteria is that the BER has to stay below $10^{-3}$ for the TD density range $[1.0, 1.1]$. That is why we switch from $6.9\%$ OP-LOCO redundancy to $17.3\%$ OT-LOCO redundancy only at $D_{\textup{TD}} = 1.1$. On the other hand, we sweep the TD channel energy metric $E_{\textup{TD}}$ in Fig.~\ref{fig_reconf2}. The two codes we use are the OP-LOCO code $\mathcal{OPC}^8_{14}$ of rate $0.8889$ and the OT-LOCO code $\mathcal{OTC}^8_{14}$ of rate $0.8125$. The reconfiguration criteria is that the BER has to stay below $5 \times 10^{-3}$ for the TD energy metric range $[400, 600]$. That is why we switch from $11.1\%$ OP-LOCO redundancy to $18.8\%$ OT-LOCO redundancy only at $E_{\textup{TD}} = 400$. Observe that the reconfiguration criteria is based on the assumption that an LDPC coding framework is part of the TDMR system to correct remaining errors.

While we are performing the reconfiguration here in a predetermined or a ``hard'' manner, one of our current research directions is about how to specify the reconfiguration point based on an online machine learning module that identifies the device status and/or an offline machine learning module that helps us reach a near-optimal compromise between storage capacity and performance. Another future research direction is about combining OP-LOCO and OT-LOCO codes effectively and efficiently with high performance modern spatially-coupled LDPC codes \cite{sy_grade} and multi-dimensional LDPC codes \cite{ahh_md} suitable for such multi-dimensional storage systems.

\begin{remark}
Additional FER/BER performance gains can be achieved if IPIS patterns where the target bit is stored on the upper or the lower tracks in each pack of $3$ down tracks are also forbidden in addition to all RTIS patterns we forbid via OT-LOCO codes. However, the capacity of a constrained code performing such a task is $0.7518$, which would result in additional storage capacity reduction compared with OT-LOCO codes. Another future research question is to investigate the need for such codes through performance analysis.
\end{remark}

\section{Simple T LOCO (ST-LOCO) Codes}\label{sec_stloco}

In this section, we discuss the advantages of constrained coding schemes for TDMR where codes are defined over an alphabet of size smaller than $8$. We then devise a new coding scheme to eliminate RTIS patterns at lower complexity and lower error propagation. We provide the mathematical analysis of this coding scheme and show what else it offers.

Since the encoding and decoding of any LOCO code are performed via the encoding-decoding rule, they become a sequence of subtractions and additions. Therefore, the encoding-decoding complexity of a LOCO code is governed by the size of the adder that executes the rule (see Table~\ref{table_3}). Typically, constrained codes defined over alphabets of higher sizes, such as OT-LOCO codes that are defined over GF$(8)$, require higher adder sizes, and thus incur relatively higher complexity.

\begin{remark}
Recall that OT-LOCO codes are optimal rate-wise and are encoded-decoded via the simple rule in \eqref{eqn_ruleot}. The phrase ``higher complexity'' for these codes in this section is always relative to other LOCO coding solutions that trade-off some rate for some complexity gain.
\end{remark}

Another aspect of constrained codes to discuss is error propagation \cite{ahh_loco}. Since LOCO codes are fixed-length codes, they do not suffer from any codeword-to-codeword error propagation. However, codeword-to-message error propagation is possible. That is, a single error in a LOCO codeword may result in multiple errors in the corresponding message upon decoding. Such error propagation does not impact FER, but it may impact message BER in case the message length is quite long.

Observe that the adder size is itself the message length. Therefore, to further reduce both complexity and error propagation while removing RTIS patterns, it is natural to think of coding solutions defined over alphabets of smaller sizes to limit the message length. In particular, we are seeking a coding solution that comprises all or some of the following: GF$(4)$ codes, GF$(2)$ codes, and uncoded setting. The cost will be a capacity/rate penalty.

There is also another perspective we considered while devising the simpler coding scheme, which is \textit{track separation}. The concept is inspired by page separation in Flash memory systems, where access speed is preserved by constrained coding solutions that do not combine all pages together in encoding-decoding \cite{ahh_rr}. In a TDMR system that adopts a wide read head, $3$ down tracks are always read together. When an OT-LOCO code is applied, all $3$ tracks have to be processed together since each column with $3$ bits is a coded GF$(8)$ symbol. Constrained coding solutions that do not adopt codes defined over GF$(8)$ can allow some tracks to be processed separately, which increases the reading speed in the TDMR system.

The idea of the proposed simple T LOCO (ST-LOCO) coding scheme, which achieves the aforementioned goals (lower complexity, lower error propagation, and track separation) in the TDMR system, is summarized in the following two steps. For each group of $3$ down tracks:
\begin{enumerate}
\item Apply a GF$(4)$ constrained code on the upper and middle tracks such that all RTIS patterns are eliminated.
\item Leave the data on the lower track uncoded.
\end{enumerate}

Let GF$(4) = \{0,1,\alpha,\alpha^2\}$ and consider the following mapping-demapping:
\begin{align}\label{eqn_gf4map}
0 &\longleftrightarrow [ 0 \textup{ } 0 ]^{\textup{T}}, \hspace{+3.6em} 1 \longleftrightarrow [ 0 \textup{ } 1 ]^{\textup{T}}, \nonumber \\
\alpha &\longleftrightarrow [ 1 \textup{ } 0 ]^{\textup{T}}, \hspace{+3.0em} \alpha^2 \longleftrightarrow [ 1 \textup{ } 1 ]^{\textup{T}}.
\end{align}
We want to eliminate all RTIS patterns in Fig.~\ref{fig:det_pattern_RTIS} by coding only on the upper and middle tracks. Therefore, RTIS patterns are eliminated by forbidding the following GF$(4)$ $2$-tuple and $3$-tuple patterns:
\begin{itemize}
\item $1$ followed by $0$ or $\alpha$, and $0$ or $\alpha$ followed by $1$.
\item $\alpha$ followed by $1$ or $\alpha^2$, and $1$ or $\alpha^2$ followed by $\alpha$.
\item $0$ or $\alpha$ followed by $1$ or $\alpha^2$ followed by $0$ or $\alpha$.
\item $1$ or $\alpha^2$ followed by $0$ or $\alpha$ followed by $1$ or $\alpha^2$.
\end{itemize}
When all of these patterns are combined to form a minimal set of first offenders, we get:
\begin{align}\label{eq:forbid_ST}
\mathcal{ST}^4 \triangleq \{01, 10, 1\alpha, \alpha1, \alpha\alpha^2, \alpha^2\alpha, 0\alpha^20, \alpha^20\alpha^2\}.
\end{align}

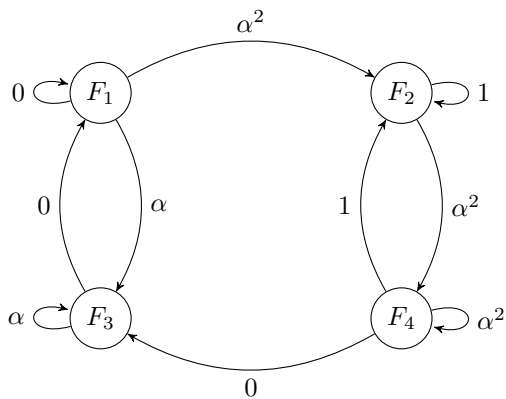
\begin{figure}[h]
\vspace{-0.5em}
\centering
\begin{tikzpicture}[->, >= stealth']
    \node [circle, draw] (one) at (-2.0, 1.5) {$F_1$};
    \node [circle, draw] (two) at (2.0, 1.5) {$F_2$};
    \node [circle, draw] (three) at (-2.0, -1.5) {$F_3$};
    \node [circle, draw] (four) at (2.0, -1.5) {$F_4$};

     \path (one) edge[loop left] node[left]{$0$} (one);
     \path (one) edge[bend left] node[ above]{$\alpha^2$} (two);
     \path (one) edge[bend left] node[right]{$\alpha$} (three);
     \path (two) edge[loop right] node[right]{$1$} (two);
     \path (two) edge[bend left] node[right]{$\alpha^2$} (four);
     \path (three) edge[loop left] node[left]{$\alpha$} (three);
     \path (three) edge[bend left] node[left]{$0$} (one);
     \path (four) edge[loop right] node[right, align=center]{$\alpha^2$} (four);
     \path (four) edge[bend left] node[left]{$1$} (two);
     \path (four) edge[bend left] node[below]{$0$} (three);
\end{tikzpicture}
\caption{An FSTD representing an infinite $\mathcal{ST}^4$-constrained sequence where patterns in $\mathcal{ST}^4$ are prevented.}
\label{fig:FSTD_ST_LOCO}
\vspace{-0.5em}
\end{figure}

The FSTD of an infinite $\mathcal{ST}^4$-constrained sequence where the patterns in $\mathcal{ST}^4$ are prevented is given in Fig.~\ref{fig:FSTD_ST_LOCO}. 
The corresponding adjacency matrix is:
\begin{equation}
    \mathbf{F}_2 = \begin{bmatrix}
        1 & 1 & 1 & 0 \\ 
        0 & 1 & 0 & 1 \\
        1 & 0 & 1 & 0 \\
        0 & 1 & 1 & 1
        \end{bmatrix}.
    \label{eq:F_ST_LOCO}
\end{equation}
The characteristic polynomial of $\mathbf{F}_2$ is:
\begin{align}
\det(x\mathbf{I}-\mathbf{F}_2) = x^4 - 4x^3 + 4x^2 - 1 = (x^2 - 2x -1) (x - 1) (x - 1).
\end{align}
Again using \eqref{eq:capacity}, the capacity $C$, in input bits per coded symbol, of ST-LOCO codes is:
\begin{align}
    C  &= \log_2(\lambda_{\textup{max}}(\mathbf{F}_2)) = \log_2(2.4142) = 1.2715.
\end{align}
Observe that $x=\lambda_{\textup{max}}(\mathbf{F}_2)$ is a root of the irreducible factor $(x^2 - 2x - 1)$ of the characteristic polynomial, which means this factor also verifies the final ST-LOCO cardinality formula \eqref{st_card}. Since coding is applied only on the upper and middle tracks while no coding is applied for the lower track, the normalized capacity $C^\textup{n}$ of the proposed ST-LOCO coding scheme is:
\begin{align}
    C^\textup{n}  &= \frac{1.2715+1}{3} = 0.7572.
\end{align}
The capacity loss compared with OT-LOCO codes is because of forbidding more patterns than needed.

For simplicity, we call a LOCO code that forbids all patterns in $\mathcal{ST}^4$ also an ST-LOCO code. An ST-LOCO code is then formally defined as follows:
\begin{definition}[ST-LOCO Code]\label{def:st_loco}
An ST-LOCO code, $\mathcal{STC}^4_{m}$, is defined by the following properties:
\begin{enumerate}
    \item Codewords in $\mathcal{STC}^4_{m}$ are defined over $\textup{GF}(4)$, the code alphabet, and are of length $m$ symbols.
    \item Codewords in $\mathcal{STC}^4_{m}$ are lexicographically ordered.
    \item Codewords in $\mathcal{STC}^4_{m}$ do not contain any patterns from the set $\mathcal{ST}^4$.
     \item Any codeword satisfying the above properties is included in $\mathcal{STC}^4_{m}$ .
\end{enumerate}
\end{definition}

After illustrating all the derivations related to OT-LOCO codes, the reader is now more familiar with the procedure we follow to design a LOCO code \cite{ahh_general}. Therefore, we will be more brief regarding the steps of designing ST-LOCO codes.

\textit{First, we specify the group structure.} Let $\zeta_1$ be in $\{0, \alpha\}$ and  $\zeta_2$ be in $\{1, \alpha^2\}$. We partition the codewords of $\mathcal{STC}^4_{m}$ according to $\mathcal{ST}^4$ into the following six final groups:
\begin{itemize}
\item {Group~1} contains all the codewords starting with $0\zeta_1$ from the left.
\item {Group~2} contains all the codewords starting with $1\zeta_2$ from the left.
\item {Group~3} contains all the codewords starting with $\alpha\zeta_1$ from the left.
\item {Group~4} contains all the codewords starting with $\alpha^2\zeta_2$ from the left.
\item {Group~5} contains all the codewords starting with $0\alpha^2\zeta_2$ from the left.
\item {Group~6} contains all the codewords starting with $\alpha^20\zeta_1$ from the left.
\end{itemize}

\textit{Second, we enumerate the codewords.} The following theorem determines how to recursively enumerate the codewords of an ST-LOCO code $\mathcal{STC}^4_{m}$.

\begin{theorem} \label{thr:step_2_st}
    The cardinality $N_4(m)$ of an ST-LOCO code $\mathcal{STC}^4_{m}$ is given by:
    \begin{equation}
        N_4(m) = 2N_4(m-1) + N_4(m-2), \textup{ } m \ge 2,
        \label{st_card}
    \end{equation}
    where the defined cardinalities are $N_4(0) \triangleq 2$ and $N_4(1) \triangleq 4$.

\begin{proof}
First of all, we note that an ST-LOCO code $\mathcal{STC}^4_{m}$ is symmetric in the sense that the number of codewords starting with $0$ ($1$) from the left equals the number of codewords starting with $\alpha^2$ ($\alpha$) from the left. We denote the cardinality of Group~$i$ in $\mathcal{STC}^4_{m}$ by $N_{4,i}(m)$.

Codewords in Group~1 in $\mathcal{STC}^4_{m}$ correspond bijectively to codewords starting with $\zeta_1$, $\zeta_1 \in \{0, \alpha\}$, from the left in $\mathcal{STC}^4_{m-1}$. Therefore and using symmetry:
\begin{equation}
N_{4,1}(m) = \frac{1}{2} N_4(m-1) = N_{4,4}(m).
\end{equation}

Codewords in Group~2 in $\mathcal{STC}^4_{m}$ correspond bijectively to codewords starting with $\zeta_2$, $\zeta_2 \in \{1, \alpha^2\}$, from the left in $\mathcal{STC}^4_{m-1}$. Therefore and using symmetry:
\begin{equation}\label{st_cardg2}
N_{4,2}(m) = \frac{1}{2} N_4(m-1) = N_{4,3}(m).
\end{equation}

Codewords in Group~5 in $\mathcal{STC}^4_{m}$ correspond bijectively to codewords starting with $\zeta_2$, $\zeta_2 \in \{1, \alpha^2\}$, from the left in $\mathcal{STC}^4_{m-2}$. Therefore and using symmetry:
\begin{equation}
N_{4,5}(m) = \frac{1}{2} N_4(m-2) = N_{4,6}(m).
\end{equation}

Collectively, we get:
\begin{equation}\label{proof_last}
N_4(m) = \sum_{i=1}^6 N_{4,i}(m) = 2N_4(m-1)+N_4(m-2).
\end{equation}
Regarding the defined cardinalities, it is natural to set $N_4(1) \triangleq 4$. Then, we know that $N_4(2) = 16-6 = 10$ directly from $\mathcal{ST}^4$. Therefore, \eqref{proof_last} for $N_4(2)$ gives $N_4(0) \triangleq 2$.
\end{proof}
\end{theorem}

\textit{Third, we determine the special cases.} The final cases derived from $\mathcal{ST}^4$ are:
\begin{itemize}
\item {Case~1.a:} $i=m-1$ and $c_i = 1$.
\item {Case~1.b:} $i=m-1$ and $c_i = \alpha$.
\item {Case~1.c:} $i=m-1$ and $c_i = \alpha^2$.
\item {Case~2:} $c_{i+1} c_i = 0\alpha$.
\item {Case~3:} $c_{i+1} c_i = 0\alpha^2$.
\item {Case~4:} $c_{i+1} c_i = 11$.
\item {Case~5:} $c_{i+1} c_i = 1\alpha^2$. 
\item {Case~6:} $c_{i+1} c_i = \alpha\alpha$. 
\item {Case~7:} $c_{i+2}c_{i+1} c_i = 0\alpha^21$. 
\item {Case~8:} $c_{i+2}c_{i+1} c_i = 0\alpha^2\alpha^2$. 
\item {Case~9:} $c_{i+1} c_i = \alpha^21$ and $c_{i+2} \neq 0$. 
\item {Case~10:} $c_{i+1} c_i = \alpha^2\alpha^2$ and $c_{i+2} \neq 0$.
\end{itemize}
Observe that Case~1, with all its subcases, is the typical case.

\textit{Fourth, we derive the symbol contribution.} We denote the contribution of $c_i$ in the case indexed by $i_\textup{c}$ by $g_{i,i_\textup{c}}(c_i)$. Let $a_i \triangleq \mathcal{L}(c_i)$, where $\mathcal{L}(c) \triangleq \text{gflog}_\alpha(c)+1$ if $c\neq0$, and $\mathcal{L}(c) \triangleq 0$ if $c=0$.

For Case~1.a, the contribution of $c_i$ is the number of codewords starting with $0$ from the left in $\mathcal{STC}^4_{i+1}$. Observe that the possible prefixes to start with are $00$, $0\alpha$, $0\alpha^21$, and $0\alpha^2\alpha^2$. Thus, the number we are seeking is the number of codewords starting with $\zeta_1$, $\zeta_1 \in \{0, \alpha\}$, in $\mathcal{STC}^4_{i}$ plus the number of codewords starting with $\zeta_2$, $\zeta_2 \in \{1, \alpha^2\}$, in $\mathcal{STC}^4_{i-1}$. This means:
\begin{equation}\label{cont_st1a}
g_{i,1.\textup{a}}(c_i) = \frac{1}{2} (N_4(i) + N_4(i-1)).
\end{equation}

For Case~1.b, the contribution of $c_i$ is the number of codewords starting with $0$ or $1$ from the left in $\mathcal{STC}^4_{i+1}$. Using code symmetry and \eqref{st_card}:
\begin{equation}\label{cont_st1b}
g_{i,1.\textup{b}}(c_i) = \frac{1}{2} N_4(i+1) = N_4(i) + \frac{1}{2} N_4(i-1).
\end{equation}

For Case~1.c, the contribution of $c_i$ is the number of codewords starting with $0$, $1$, or $\alpha$ from the left in $\mathcal{STC}^4_{i+1}$. Using \eqref{cont_st1b} and \eqref{st_cardg2} gives:
\begin{equation}\label{cont_st1c}
g_{i,1.\textup{c}}(c_i) = \frac{1}{2} (3N_4(i) + N_4(i-1)).
\end{equation}

For Case~2, the contribution of $c_i$ is the number of codewords starting with $00$ from the left in $\mathcal{STC}^4_{i+2}$, which is the number of codewords starting with $0$ from the left in $\mathcal{STC}^4_{i+1}$. From \eqref{cont_st1a}:
\begin{equation}\label{cont_st2}
g_{i,2}(c_i) = \frac{1}{2} (N_4(i) + N_4(i-1)).
\end{equation}

For Case~3, the contribution of $c_i$ is the number of codewords starting with $0\zeta_1$, $\zeta_1 \in \{0, \alpha\}$, from the left in $\mathcal{STC}^4_{i+2}$, which is the number of codewords starting with $\zeta_1$ from the left in $\mathcal{STC}^4_{i+1}$. Therefore,
\begin{equation}\label{cont_st3}
g_{i,3}(c_i) = \frac{1}{2} N_4(i+1) = N_4(i) + \frac{1}{2} N_4(i-1).
\end{equation}

For Case~4, the contribution of $c_i$ is the number of codewords starting with $10$ from the left in $\mathcal{STC}^4_{i+2}$. However, $10$ is a forbidden pattern. Therefore,
\begin{equation}\label{cont_st4}
g_{i,4}(c_i) = 0.
\end{equation}

For Case~5, the contribution of $c_i$ is the number of codewords starting with $11$ from the left in $\mathcal{STC}^4_{i+2}$, which is the number of codewords starting with $1$ from the left in $\mathcal{STC}^4_{i+1}$. From \eqref{st_cardg2}:
\begin{equation}\label{cont_st5}
g_{i,5}(c_i) = \frac{1}{2} N_4(i).
\end{equation}

For Case~6, the contribution of $c_i$ is the number of codewords starting with $\alpha0$ from the left in $\mathcal{STC}^4_{i+2}$, which is the number of codewords starting with $0$ from the left in $\mathcal{STC}^4_{i+1}$. From \eqref{cont_st1a}:
\begin{equation}\label{cont_st6}
g_{i,6}(c_i) = \frac{1}{2} (N_4(i) + N_4(i-1)).
\end{equation}

For Case~7, the contribution of $c_i$ is the number of codewords starting with $0\alpha^20$ from the left in $\mathcal{STC}^4_{i+3}$. However, $0\alpha^20$ is a forbidden pattern. Therefore,
\begin{equation}\label{cont_st7}
g_{i,7}(c_i) = 0.
\end{equation}

For Case~8, the contribution of $c_i$ is the number of codewords starting with $0\alpha^21$ from the left in $\mathcal{STC}^4_{i+3}$, which is the number of codewords starting with $\zeta_2$, $\zeta_2 \in \{1, \alpha^2\}$, from the left in $\mathcal{STC}^4_{i}$. Therefore,
\begin{equation}\label{cont_st8}
g_{i,8}(c_i) = \frac{1}{2} N_4(i).
\end{equation}

For Case~9, the contribution of $c_i$ is the number of codewords starting with $\alpha^20$ from the left in $\mathcal{STC}^4_{i+2}$, which is the number of codewords starting with $\zeta_1$, $\zeta_1 \in \{0, \alpha\}$, from the left in $\mathcal{STC}^4_{i}$. Therefore,
\begin{equation}\label{cont_st9}
g_{i,9}(c_i) = \frac{1}{2} N_4(i).
\end{equation}

For Case~10, the contribution of $c_i$ is the number of codewords starting with $\alpha^20\zeta_1$, $\zeta_1 \in \{0, \alpha\}$, or $\alpha^21\zeta_2$, $\zeta_2 \in \{1, \alpha^2\}$, from the left in $\mathcal{STC}^4_{i+2}$, which is the number of all codewords in $\mathcal{STC}^4_{i}$. Therefore,
\begin{equation}\label{cont_st10}
g_{i,10}(c_i) = N_4(i).
\end{equation}

\textit{Fifth, we formulate the encoding-decoding rule.} The following theorem states a bijective index-codeword relation, which is the base of ST-LOCO encoding-decoding algorithms.
\begin{theorem}\label{thr:step_5st}
Let $\mathbf{c}$ be an ST-LOCO codeword in $\mathcal{STC}^4_{m}$. The relation between the lexicographic index $g(\mathbf{c})$ of this codeword and the codeword itself is given by:    \begin{align}\label{eqn_rulest}
g(\mathbf{c}) = \sum_{i=0}^{m-1} \left [ \left ( \frac{1}{2} \theta_{i,1} + \theta_{i,2} \right ) N_4(i) + \frac{1}{2} \theta_{i,3} N_4(i-1) \right ],
\end{align}
where $\theta_{i,1} = y_{i,1} + y_{i,3} + y_{i,4}$, $\theta_{i,2} = y_{i,2} + y_{i,3} + y_{i,5}$, and $\theta_{i,3} = y_{i,1} + y_{i,2} + y_{i,3}$. Moreover, \\
\indent $y_{i,1}=1$ if $c_i = c_{m-1} = 1$ or $c_{i+1}c_i = 0\alpha$ or $c_{i+1}c_i = \alpha\alpha$ (Case 1.a or 2 or 6), and $y_{i,1}=0$ otherwise,\\
\indent $y_{i,2}=1$ if $c_i = c_{m-1} = \alpha$ or $c_{i+1}c_i = 0\alpha^2$ (Case 1.b or 3), and $y_{i,2}=0$ otherwise,\\
\indent $y_{i,3}=1$ if $c_i = c_{m-1} = \alpha^2$ (Case 1.c), and $y_{i,3}=0$ otherwise,\\
\indent $y_{i,4}=1$ if $c_{i+1}c_i = 1\alpha^2$ or $c_{i+2}c_{i+1}c_i = 0\alpha^2\alpha^2$ or $c_{i+1}c_i = \alpha^21$ while $c_{i+2} \neq 0$ (Case 5 or 8 or 9), and $y_{i,4}=0$ otherwise,\\
\indent $y_{i,5}=1$ if $c_{i+1}c_i = \alpha^2\alpha^2$ while $c_{i+2} \neq 0$ (Case 10), and $y_{i,5}=0$ otherwise.

\begin{proof}
We have already computed symbol contributions for all final cases in the fourth step above. Now, we will merge them into one relation for $g_i(c_i)$. Since every non-zero contribution in the equations \eqref{cont_st1a}--\eqref{cont_st10} is function of $N_4(i)$ and/or $N_4(i-1)$, we need two merging functions $f_1^{\textup{mer}}(\cdot)$ and $f_2^{\textup{mer}}(\cdot)$, respectively. That is:
\begin{equation}\label{eqn_mergest}
g_i(c_i) = f_1^{\textup{mer}}(\cdot)N_4(i) + f_2^{\textup{mer}}(\cdot)N_4(i-1).
\end{equation}

Using the definition of $y_{i,j}$, for all $j \in \{1, 2, \dots, 5\}$, $\theta_{i,k}$, for all $k \in \{1, 2, 3\}$, as well as \eqref{cont_st1a}--\eqref{cont_st10}, the first merging function can be written as follows:
\begin{align}\label{eqn_f1st}
f_1^{\textup{mer}}(\cdot) &= \frac{1}{2}y_{i,1} + y_{i,2} + \frac{3}{2}y_{i,3} + \frac{1}{2}y_{i,4} + y_{i,5} \nonumber \\ &= \frac{1}{2}\theta_{i,1} + \theta_{i,2}.
\end{align}
Moreover, using the definition of $y_{i,j}$, for all $j \in \{1, 2, \dots, 5\}$, $\theta_{i,k}$, for all $k \in \{1, 2, 3\}$, as well as \eqref{cont_st1a}--\eqref{cont_st10}, the second merging function can be written as follows:
\begin{align}\label{eqn_f2st}
f_2^{\textup{mer}}(\cdot) &= \frac{1}{2}y_{i,1} + \frac{1}{2}y_{i,2} + \frac{1}{2}y_{i,3} \nonumber \\ &= \frac{1}{2}\theta_{i,3}.
\end{align}

Substituting \eqref{eqn_f1st} and \eqref{eqn_f2st} in \eqref{eqn_mergest} leads to:
\begin{align}
g(\mathbf{c}) &= \sum_{i=0}^{m-1} g_i(c_i) = \sum_{i=0}^{m-1} \left [ f_1^{\textup{mer}}(\cdot)N_4(i) + f_2^{\textup{mer}}(\cdot)N_4(i-1) \right ] \nonumber \\ &= \sum_{i=0}^{m-1} \left [ \left ( \frac{1}{2} \theta_{i,1} + \theta_{i,2} \right ) N_4(i) + \frac{1}{2} \theta_{i,3} N_4(i-1) \right ],
\end{align}
which completes the proof of the encoding-decoding rule of ST-LOCO codes.
\end{proof}
\end{theorem}

\begin{example}
Consider the ST-LOCO code $\mathcal{STC}^4_{5}$. Using Theorem~\ref{thr:step_2_st},
\begin{equation}
N_4(0) \triangleq 2, \textup{ } N_4(1) \triangleq 4, \textup{ } N_4(2) = 10, \textup{ } N_4(3) = 24, N_4(4) = 58, \text{ and } N_4(5) = 140.
\end{equation}
Suppose that we want to compute the index of the codeword $\mathbf{c} = \alpha^2\alpha^2\alpha^2\alpha^2\alpha^2$. Observe that this codeword has to be the last codeword in $\mathcal{STC}^4_{5}$ according to the lexicographic order. Therefore, we already know that:
\begin{equation}\label{known_st}
g(\mathbf{c} = \alpha^2\alpha^2\alpha^2\alpha^2\alpha^2) = N_4(5)-1 = 139.
\end{equation}
Now, we verify this via the encoding-decoding rule of ST-LOCO codes. Using Theorem~\ref{thr:step_5st} we conclude the following:
\begin{itemize}
\item For $c_4$, $y_{4,3} = 1$. Thus, $\theta_{4,1} = \theta_{4,2} = \theta_{4,3} = 1$.
\item For $c_i$, $i \in \{0,1,2,3\}$, $y_{i,5} = 1$. Thus, $\theta_{i,1} = \theta_{i,3} = 0$ and  $\theta_{i,2} = 1$, for all $i \in \{0,1,2,3\}$.
\end{itemize}
Next, we directly substitute in \eqref{eqn_rulest} to compute the index:
\begin{align}
g(\mathbf{c} = \alpha^2\alpha^2\alpha^2\alpha^2\alpha^2) &= \left [ \frac{3}{2}N_4(4) + \frac{1}{2}N_4(3)  \right ] + N_4(3) + N_4(2) + N_4(1) + N_4(0) \nonumber \\ &= 
[87+12] + 24 + 10 + 4 + 2 = 139,
\end{align}
which is perfectly consistent with what we know in this case from \eqref{known_st}.
\end{example}

As usual, we will not discuss the \textit{sixth step} in detail. Instead, we will discuss bridging, self-clocking, and finite-length rates of ST-LOCO codes.

\textbf{ST-LOCO bridging:} Recall the set of forbidden patterns:
\begin{align}
\mathcal{ST}^4 \triangleq \{01, 10, 1\alpha, \alpha1, \alpha\alpha^2, \alpha^2\alpha, 0\alpha^20, \alpha^20\alpha^2\}. \nonumber
\end{align}
For the sake of brevity, we again use the following format when studying the bridging between two consecutive ST-LOCO codewords in $\mathcal{STC}^4_{m}$ ``right-most symbols of codeword at $t$ -- bridging pattern -- left-most symbols of codeword at $t+1$''. Denote the bridging sequence by $\mathbf{d}$, whose fixed length is unspecified yet.

Consider the following case:
\begin{equation}\label{eqn_stcase}
1 - \mathbf{d} - \alpha.
\end{equation}
We first examine the situation of $\mathbf{d}$ of length $1$. Because of the right-most symbol of the codeword at $t$, $\mathbf{d}$ cannot be in $\{0, \alpha\}$. Moreover, because of the left-most symbol of the codeword at $t+1$, $\mathbf{d}$ cannot be in $\{1, \alpha^2\}$. Thus, $\mathbf{d}$ cannot be of length $1$. Next, we examine the situation of $\mathbf{d}$ of length $2$, i.e., $\mathbf{d} = d_1 d_0$. By examining all possible $16$ bridging patterns, we can conclude that the only possible bridging pattern of length $2$, which does not create any pattern in $\mathcal{ST}^4$, is:
\begin{align}\label{eqn_stbrdg_1}
\mathbf{d} = d_1d_0 = \alpha^2 0.
\end{align}
Note that such bridging will result in encoding $\log_2 1 = 0$ input message bits within the bridging interval. 

Because of the above analysis, we move on to the situation of $\mathbf{d}$ of length $3$, i.e., $\mathbf{d} = d_2 d_1 d_0$. We are still studying the case in \eqref{eqn_stcase}. Because of the right-most symbol of the codeword at $t$, $d_2$ has to be in $\{1, \alpha^2\}$. Moreover, because of the left-most symbol of the codeword at $t+1$, $d_0$ has to be in $\{0, \alpha\}$. Given the options for $d_2$ and $d_0$, $d_1$ has to be in $\{0, \alpha^2\}$. While this implies $8$ options for $\mathbf{d} = d_2 d_1 d_0$, not all of them are possible to use. For example, we cannot use $\mathbf{d} = 100$. However, all
\begin{equation}\label{eqn_stbrdg_1}
\mathbf{d} = d_2 d_1 d_0 \in \{1\alpha^20, \alpha^200, \alpha^20\alpha, \alpha^2\alpha^20\}
\end{equation}
are possible to use as bridging patterns.

By checking all the possible cases for the right-most symbols of the codeword at $t$ and the left-most symbols of the codeword at $t+1$, one can conclude that the aforementioned case results in the highest level of restrictions on $3$-tuple $\mathbf{d}$; that is, $\mathbf{d}$ has only $4$ possible options. We abide by these restrictions in all cases and thus, we specify all bridging patterns to use based on the rules:
\begin{enumerate}
\item Each case has at least $4$ possible bridging patterns out of which, we will use only $4$.
\item Bridging patterns must prevent forbidden patterns from appearing at the transition between codewords.
\item Each bridging pattern $\mathbf{d}$ is of length $3$, i.e., $\mathbf{d} = d_2d_1d_0$.
\end{enumerate}

For brevity, we skip listing all bridging patterns for all cases. Moreover, in ST-LOCO bridging, we again use bridging symbols to encode binary input message bits in order to increase the finite-length rate. In particular, since we use $4$ possible bridging patterns for every case, we encode $2$ additional input message bits via these bridging symbols. It is important to highlight that the impact of such a simple idea on the finite-length rate is significant.

\textbf{ST-LOCO self-clocking:} In order to achieve self-clocking, very long same-symbol sequences should not be allowed. We do not need to eliminate any codewords to achieve that here. The reason is that our aforementioned bridging is devised such that a transition from a symbol to a different symbol occurs within the bridging interval or/and directly before/after the bridging interval. This makes our ST-LOCO codes \textit{intrinsically self-clocked}. Denote the maximum number of consecutive GF$(4)$ symbols ($2$-bit columns) that are identical in an ST-LOCO stream coded via $\mathcal{STC}^4_{m}$ and stored in a TDMR device by $k_\textup{eff}^\textup{st}$. Then,
\begin{equation}
k_\textup{eff}^\textup{st} = m+4.
\end{equation}
Observe that this self-clocking discussion applies as is for the whole group of $3$ down tracks since the data on the lower track in the ST-LOCO coding scheme is random (uncoded) and cannot be controlled.

\textbf{ST-LOCO rates:} 
The rate of an ST-LOCO code $\mathcal{STC}^4_{m}$ is the number of message bits divided by the number of coded symbols. We have $s = \left \lfloor \log_2 (N_4(m)) \right \rfloor$ message bits converted into an ST-LOCO codeword and $2$ message bits encoded within bridging. Observe that $s$ is also the adder size, and therefore, it dictates the complexity of encoding-decoding. Moreover, we have $m+3$ coded symbols. Therefore, the rate of an ST-LOCO code $\mathcal{STC}^4_{m}$ is:
\begin{equation}\label{eqn_rate_st}
R_\textup{ST-LOCO} = \frac{s+2}{m+3} = \frac{\left \lfloor \log_2 (N_4(m)) \right \rfloor +2}{m+3}.
\end{equation}
Observe that what matters more is the normalized rate of the ST-LOCO coding scheme, i.e., the rate including the uncoded lower track data. The normalized rate of an ST-LOCO coding scheme is:
\begin{equation}\label{eqn_raten_st}
R_\textup{ST-LOCO}^\textup{n} = \frac{1}{3}\left [ \frac{s+2}{m+3} + 1 \right ] = \frac{1}{3}\left [ \frac{\left \lfloor \log_2 (N_4(m)) \right \rfloor+2}{m+3} + 1 \right ].
\end{equation}
We also note that our ST-LOCO coding scheme is \textit{capacity-achieving}.

\begin{table}
\caption{Normalized Rates and Adder Sizes of the ST-LOCO Coding Scheme Comprising $\mathcal{STC}^4_m$ for Different Values of $m$. The Normalized Capacity Is $0.7572$.}
\vspace{-0.5em}
\centering
\scalebox{1.00}
{
\begin{tabular}{|c|c|c|}
\hline
\makecell{$m$} & \makecell{$R_{\textup{ST-LOCO}}^{\textup{n}}$}  & \makecell{Adder size} \\
\hline
$5$ & $0.7083$ & $7$ bits \\
\hline
$9$ & $0.7222$ & $12$ bits \\
\hline
$12$ & $0.7333$ & $16$ bits \\
\hline
$23$ & $0.7436$ & $30$ bits \\
\hline
$34$ & $0.7477$ & $44$ bits \\
\hline
$49$ & $0.7500$ & $63$ bits \\
\hline
\end{tabular}}
\label{table_4}
\vspace{-0.5em}
\end{table}

Table~\ref{table_4} gives the normalized rates and adder sizes of the ST-LOCO coding scheme for different values of $m$. Observe how finite-length rates approach capacity. More importantly, observe that capacity-approaching rates can be achieved with quite small adder sizes. In particular, we can achieve a normalized rate within $6.46\%$ ($4.62\%$ and $3.16\%$) from the normalized capacity using an adder of size only $7$ ($12$ and $16$) bits. These quite small adder sizes result in remarkably lower complexity and remarkably lower error propagation on the upper and middle down tracks. Recall that in our ST-LOCO coding scheme, there does not exist any error propagation on the lower track in each group of $3$ down tracks.

\section{Conclusion}\label{sec_conc}

We experimentally showed that as the media noise impact increases because of high TD density and/or low TD energy, new error-prone data patterns, namely IPIS patterns, emerge in TDMR systems. We characterized the set of all error-prone patterns in TDMR, namely RTIS patterns, and we showed that the capacity of a constrained code that forbids all RTIS patterns is surprisingly high. We analyzed and designed constrained codes, namely OT-LOCO codes, that forbid all RTIS patterns. In particular, we derived a recursive relation that gives the cardinality of OT-LOCO codes. Moreover, we devised a simple encoding-decoding rule for these codes, which relates each OT-LOCO codeword to its lexicographic index bijectively. This encoding-decoding rule allows simple reconfiguration of the proposed codes. We introduced a novel bridging technique for these codes and discussed their capacity achievability. We presented simulation results that demonstrate remarkable performance and density/energy gains achieved by OT-LOCO codes in TDMR systems compared with both OP-LOCO codes and the uncoded setting. We also showed how to reconfigure the constrained code to further preserve storage capacity while having a strict performance criterion. We proposed ST-LOCO coding to remove RTIS patterns at lower complexity and error propagation, but with some rate penalty, and showed how ST-LOCO codes can increase reading speed. We suggest that OT-LOCO codes and the concept of reconfiguration could help the evolution of the TDMR technology.

\section*{Acknowledgment}

We would like to thank Mohsen Bahrami, Prof. Bane Vasic, and also Beyza Dabak for providing and developing the TDMR system model that we updated and used to generate the results in Section~\ref{sec_motiv} and Section~\ref{sec_sims}.

\ifCLASSOPTIONcaptionsoff
  \newpage
\fi

\end{document}